\documentclass[twocolumn]{aastex63}
\usepackage{color}
\usepackage[titletoc]{appendix}
\usepackage[fleqn]{amsmath}
\usepackage{amssymb}
\usepackage{mathtools}
\usepackage{float}
\usepackage{comment}
\usepackage{enumitem}
\usepackage{natbib}
\usepackage{graphicx}
\usepackage{bm}
\usepackage{totcount}

\shortauthors{Spalding \& Millholland}
\shorttitle{Star vs. Giant}

\begin{document} 

\title{Stellar Oblateness versus Distant Giants in Exciting \textit{Kepler} Planet Mutual Inclinations}
\author{Christopher Spalding}
\altaffiliation{51 Pegasi b Fellow}
\affiliation{Department of Astronomy, Yale University, New Haven, CT 06511, USA}

\author{Sarah C. Millholland}
\altaffiliation{NASA Sagan Fellow}
\affiliation{Department of Astronomy, Yale University, New Haven, CT 06511, USA}
\affiliation{Department of Astrophysical Sciences, Princeton University, Princeton, NJ 08544, USA}
\email{christopher.spalding@yale.edu}

\begin{abstract}
An overabundance of single-transiting \textit{Kepler} planets suggests the existence of a sub-population of intrinsically multi-planet systems possessing large mutual inclinations. However, the origin of these mutual inclinations remains unknown. Recent work has demonstrated that mutual inclinations can be excited soon after protoplanetary disk-dispersal due to the oblateness of the rapidly-rotating host star, provided the star is tilted. Alternatively, distant giant planets, which are common in systems of close-in \textit{Kepler} planets, could drive up mutual inclinations. The relative importance of each of these mechanisms has not been investigated. Here, we show that the influence of the stellar oblateness typically exceeds that of an exterior giant soon after planet formation. However, the magnitude of the resulting mutual inclinations depends critically upon the timescale over which the natal disk disperses. Specifically, we find that if the disk vanishes over a timescale shorter than $\sim 10^{3-4}$ years, comparable to the viscous timescale of the inner $\sim0.2$\,AU, the inner planets impulsively acquire misalignments that scale with the stellar obliquity. In contrast, if the disk disperses slowly, the inner planets remain coplanar. They first align with the stellar equator but subsequently realign with the distant giant's plane as the star spins down. Our findings are consistent with recent observations that giants tend to be aligned with close-in multis but misaligned with singles. Stellar obliquity measurements offer a promising test of our proposed framework.\\
\end{abstract}


\section{Introduction}

Within our Solar System, the planetary orbits share a common plane to within a few degrees. This copanarity was inherited from a protoplanetary disk \citep{Kant1755,Laplace1796} and was retained throughout 4.5 Gyr of our Solar System's history \citep{kenyon2004stellar,juric2008dynamical,johansen2012can,ford2008origins,li2015cross}. Extrasolar planetary systems likewise form within thin disks, but it remains unknown what proportion of these systems retain their primordial coplanarity throughout their lifetimes \citep{fabrycky2014architecture,batalha2013planetary,burke2015terrestrial}.

Unfortunately, except for limited special cases \citep[e.g.][]{laughlin2002dynamical, 2017AJ....153...45M}, mutual inclinations are typically difficult to measure \citep{xie2016exoplanet}. Within systems already-known to possess multiple transiting planets, mutual inclinations tend to be small, about $\sim1-2^{\circ}$ \citep{fabrycky2014architecture}, increasing closer to the star \citep{dai2018larger}. However, such approaches only serve as lower limits, given the limitation that multiple planets must be fairly close to coplanarity in order to appear in transit.  

At the population level, larger mutual inclinations tend to increase the number of single-transiting planets observed relative to multi-transiting systems \citep{lissauer2011architecture,becker2016oscillations,zink2019accounting}. Comparisons of transit multiplicities repeatedly find an over-abundance of transiting singles, inconsistent with uniformly low mutual inclinations among intrinsically multi-planet systems \citep{tremaine2012statistics,johansen2012can,ballard2016kepler,zhu201830,2019MNRAS.490.4575H}. Accordingly, either there exists a separate population of intrinsically single planetary systems, or mutual inclinations large enough to reduce the transit number are widespread. The latter of these explanations is generally favored \citep{zhu201830,2019MNRAS.490.4575H}, i.e., that some mechanism dynamically heats a substantial fraction of close-in \textit{Kepler} systems, raising their mutual inclinations.

The dominant origin of mutual inclinations within close-in planetary systems has remained uncertain. A recently-proposed misalignment mechanism is the quadrupolar gravitational potential of a tilted, rapidly-rotating host star \citep{spalding2016spin,spalding2018resilience,li2020mutual}. Specifically, during the disk-hosting stage, stars typically exhibit short rotation periods of several days (driving significant oblateness) and distended radii \citep{bouvier2014angular,shu1987star}. Combined, these physical attributes greatly enhance the gravitational quadrupolar moment of the star, which forces close-in planetary orbital planes to rapidly precess about the stellar spin axis \citep{murray1999solar,spalding2018resilience}. The precession rate is faster at shorter orbital periods, such that each planetary orbit precesses at a different rate, driving their planes apart. For sufficiently large stellar obliquities, this mechanism generates mutual inclinations between close-in planets that are sufficient to reduce the number of planets observed in transit \citep{spalding2016spin,li2020mutual}. 

Significant, non-zero stellar obliquities have been measured in stars hosting a wide range of planets, from single hot Jupiters to multiple super-Earths \citep{winn2010hot,albrecht2012obliquities,dai2017oblique,2017AJ....154..270W,yee2018hat,wang2018stellar,dalal2019nearly,kamiaka2019misaligned}. Though the origin of stellar obliquities remains unclear\footnote{It is likely that some fraction of these obliquities arose after the disk-hosting stage due to dynamical interactions such as the Lidov-Kozai effect \citep{ngo2016friends,naoz2016eccentric} and/or planet-planet scattering  \citep{chatterjee2008dynamical}. However, only those excited prior to disk dispersal are relevant to our discussion here.}, there have been numerous proposed pathways toward the excitation of spin-orbit misalignments during the disk-hosting stage, including perturbations upon the natal disk from stellar companions \citep{batygin2012primordial,spalding2014early,zanazzi2018planet}, star-disk magnetic torques \citep{spalding2015magnetic,lai2014star}, or turbulence during star formation \citep{bate2010chaotic,fielding2015turbulent,spalding2014alignment}. Moreover, a limited but growing number of disk-hosting stars have been observed to possess non-zero obliquities \citep{davies2019star}, indicating the potential for a widespread existence of star-disk misalignments.

An additional mechanism for the generation of mutual inclinations among close-in planets is the secular gravitational influence of an inclined, distant giant planet \citep{hansen2017perturbation,2017AJ....153...42L,2018MNRAS.478..197P}. Similarly to the stellar quadrupole, the giant's influence forces the inner planetary orbits to differentially precess, driving the planets away from coplanarity \citep{2018MNRAS.478..197P}. Recently, radial velocity surveys have revealed that over $\sim 30\%$ of already known close-in planetary systems possess a distant giant companion planet, i.e., one similar in mass to Jupiter, orbiting between $\sim 1-20$\,AU  \citep{bryan2016statistics,2019AJ....157...52B,kawahara2019transiting,fernandes2019hints}. This significantly exceeds the $\sim6\%$ of field stars hosting such giant planets \citep{cumming2008keck,wittenmyer2016anglo}, suggesting a correlation between close-in super Earths and distant giants \citep{2018AJ....156...92Z}.

Exterior giant planets typically excite mutual inclinations only if they themselves are inclined with respect to an interior system of smaller planets \citep{2017AJ....153...42L,2018MNRAS.478..197P}. Several examples exist of giant planets possessing mutual inclinations, such as Kepler-108 ($\sim 24^\circ$ mutual inclination; \citealt{mills2017kepler}) and Kepler-419 (mutual inclination of $\sim9^\circ$; \citealt{dawson2014large}), together with indirect signatures of exterior giants inclined to warm Jupiters \citep{dawson2014class}. However, distant giants are expected to form in the same plane as their natal disks, such that the requisite inclination of the exterior giant must arise through some dynamical process, such as planet-planet scattering \citep{rasio1996dynamical,gratia2017outer}, secular chaos \citep{wu2011secular}, or perturbations from a binary companion \citep{wu2003planet}, occurring subsequent to disk-dispersal. 

In order provide empirical constraints upon the inclinations of distant giant planets, \citet{masuda2020mutual} searched for transits of giant planets in systems known to host close-in transiting planets. The number of giants observed to transit in known multi-transiting systems suggested that distant giants in these systems are typically well-aligned to within $\sim4^{\circ}$. In contrast, no giants were found in already-known transit singles, hinting at a larger spread of mutual inclinations among singles. 

Such an absence of giant planets found to co-transit with singles is consistent with the idea that dynamical excitation arises from a population of inclined, distant giants \citep{2017AJ....153...42L}. However, an equally consistent explanation of the observations is that the stellar quadrupole misaligned the inner system of planets \citep{spalding2016spin,li2020mutual}, such that the inner planets' inclinations relative to the giant's plane arose as a consequence. Currently, these two scenarios are difficult to distinguish empirically. Moreover, it is important to note that both mechanisms may be occurring simultaneously, with varying prominence from system to system, but a general study of their relative significance has yet to be performed.

In this work, we explore the relative capabilities of the stellar oblateness and distant giants to drive mutual inclinations within close-in planetary systems. Both mechanisms require additional processes to tilt the star and/or incline the giant's orbit. However, we show that in general, the rapid-rotation of young stars leads to a stronger secular perturbation upon close-in planetary systems than does a giant outside of 2\,AU. Importantly, we find that in order for the stellar quadrupole to induce mutual inclinations, the natal disk must disperse sufficiently rapidly, such that the planets are impulsively driven to a state of spin-orbit misalignment after disk dispersal. In this case, the stellar obliquity readily misaligns the planetary obits. 

In the contrasting case of a slowly dispersing disk, the inner planets are able to adiabatically realign with the stellar spin axis, maintaining their primordial orbital coplanarity in the process. If a distant giant is present, the subsequent spin down of the host star removes the stellar quadrupole, causing the close-in planets to adiabatically align with the giant's plane. For a pictorial summary of our main conclusions, see Figure~\ref{schematic}.

The paper is organized as follows. We begin by describing the secular theory used to simulate the interaction between the star, planets and disk (Section \ref{sec: physical set-up}). Next, we compare the secular influences of the giant and stellar quadrupole (Section~\ref{sec: Compare}), and simulate their influences upon inclined planetary systems (Section~\ref{sec: Dependence}). In Section~\ref{sec: disk-driven initial conditions}, we study the disk's role in setting the initial conditions of the system. In Section~\ref{sec: Nbody} we perform full $N$-body simulations to validate and extend our secular investigations, before discussing the implications of our results in Section~\ref{sec: discussion}. We conclude in Section~\ref{sec: summary}.

\section{Physical set-up}
\label{sec: physical set-up}

\begin{figure}[!ht]
\centering
\includegraphics[trim=0cm 0cm 0cm 0cm, clip=true,width=1\columnwidth]{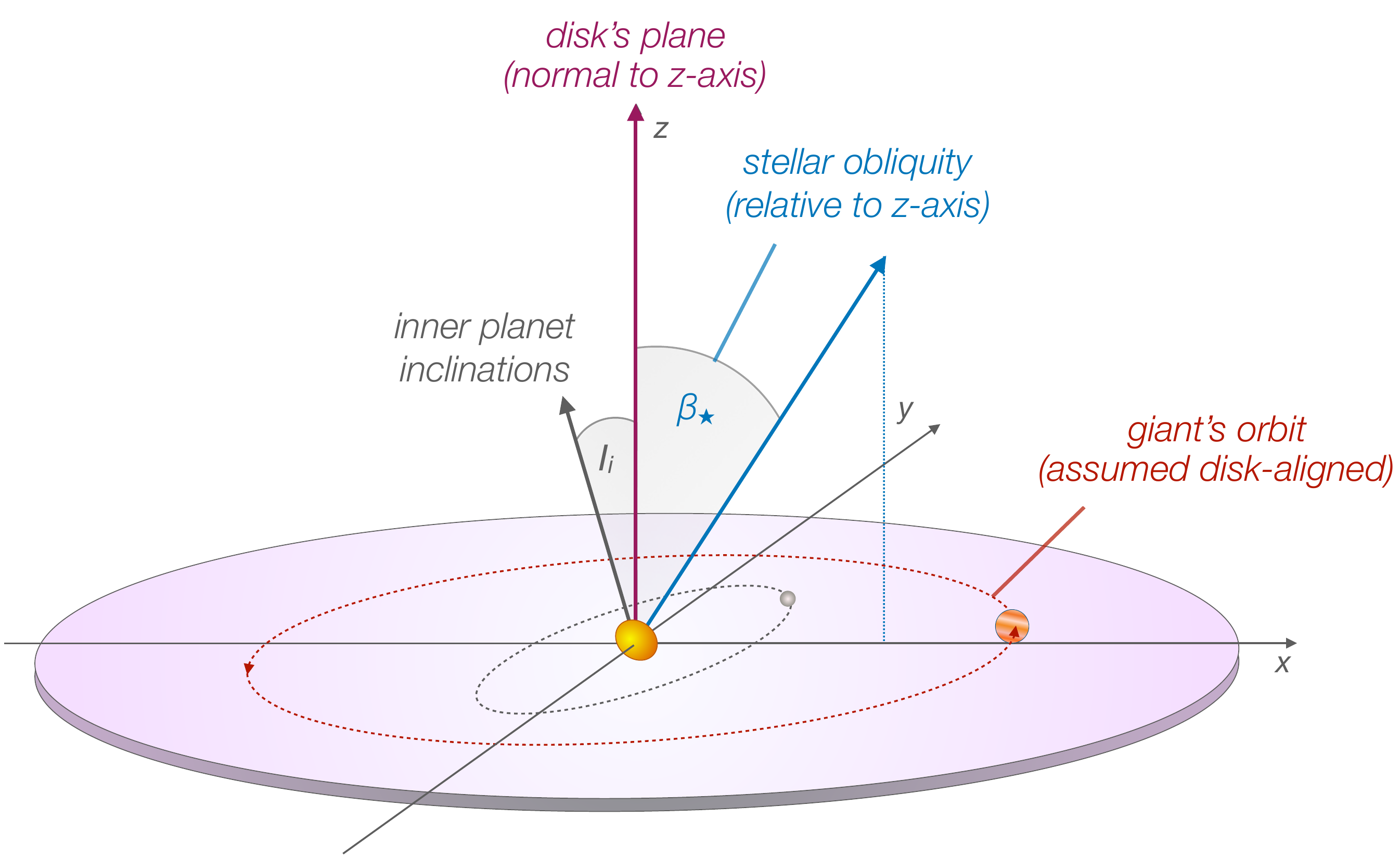}
\caption{Physical set-up of the problem. A system of $n_p$ close-in planets orbits around a star with quadrupole moment $J_2$ and rotation axis (blue arrow) tilted by $\beta_\star$ from the $z$-axis. A giant planet orbits at a semi-major axis $a_G$ exterior to the system within the $x-y$ plane, which is assumed to coincide with the natal protoplanetary disk's plane. Each inner planet possesses an orbital inclination defined by an angle $I_i$ between the orbit normal and the $z$-axis. }\label{Simple_Schem}
\end{figure}

Let us suppose that a system of $n_p$ close-in planets emerge from their protoplanetary disk with circular orbits, semi-major axes $a_i\lesssim0.5$\,AU and masses $m_i$. Exterior to this close-in system orbits a giant planet of mass $m_G\gg m_i$ and semi-major axis $a_G\gg a_i$ (see Figure~\ref{Simple_Schem}). The system orbits a host star possessing a second gravitational (quadrupole) moment $J_2$, where larger values of $J_2$ correspond to a more oblate and rapidly-rotating star. Throughout this work, we define the stellar obliquity, labelled $\beta_\star$, as the angle between the stellar spin vector and the $z$-axis. Likewise, the orbit normal vectors of each planet are inclined relative to the $z$-axis by an amount $I_i$. Our goal is to compute the time evolution of the inclinations of the inner $n_p$ planets.

\label{schematic_alt}

\subsection{2-planet secular model}
 
In order to describe our mathematical approach, we begin with $n_p=2$ inner planets, but generalize the method in the next section. We assume that the planets' orbital periods are far from integer ratios, allowing the use of secular dynamics, whereby each orbit is effectively approximated as a massive wire \citep{morbidelli2002modern}. Furthermore, we restrict attention to small inclinations, such that Laplace-Lagrange secular theory is appropriate \citep{murray1999solar}. 

The interior planets are influenced by secular perturbations from the exterior giant, the host star's quadrupole, and the other close-in planets. The ratio of the stellar angular momentum to the orbital angular momentum of planet $p$ is given by
\begin{align}
    \frac{l_\star M_\star \omega_\star R_\star^2}{m_p\sqrt{G M_\star a_p}}&\sim 10 \bigg(\frac{M_\star}{M_\odot}\bigg)^{\frac{1}{2}}\bigg(\frac{R_\star}{R_\odot}\bigg)^2\bigg(\frac{P_\star}{10\textrm{day}}\bigg)^{-1}\nonumber\\
    &\times\bigg(\frac{a_p}{0.3AU}\bigg)^{-\frac{1}{2}}\bigg(\frac{m_p}{10M_\oplus}\bigg)^{-1},
\end{align}
where we define the stellar mass, $M_\star$, radius, $R_\star$, spin rate, $\omega_\star=2\pi/P_\star$, and moment of inertia factor, $l_\star=0.2$ \citep{batygin2013magnetic}. This ratio is much greater than unity within the regime we consider. Similarly, the distant giant's orbital angular momentum greatly exceeds that of the inner planetary system. Thus, we assume that both the giant's orbit and the stellar orientation are fixed throughout the secular calculation (an assumption that is relaxed in Section~\ref{sec: Nbody}). 

The evolution of the system is conveniently described by Hamiltonian dynamics in terms of the canonical Poincar\'e variables \citep{morbidelli2002modern}:
\begin{align}
Z_i&\equiv m_i\sqrt{GM_\star a_i}\big[1-\cos(I_i)\big]\approx\frac{I_i^2}{2}m_i\sqrt{GM_\star a_i}\nonumber\\
\Lambda_i&\equiv m_i\sqrt{GM_\star a_i}\nonumber\\
z_i&\equiv -\Omega_i,
\end{align}
where $\Omega_i$ are the longitudes of ascending node.

Without loss of generality, we set $\Omega_\star=0$, and $Z_G=0$. In the case thus described, we may write the Laplace-Lagrange Hamiltonian to lowest order as \citep{murray1999solar},
\begin{align}
\mathcal{H}&=\frac{3}{2}J_2\bigg(\frac{R_\star}{a_1}\bigg)^2\frac{GMm_1}{a_1}\bigg[\frac{Z_1}{\Lambda_1}-\beta_\star \sqrt{\frac{2Z_1}{\Lambda_1}}\cos(z_1)\bigg]\nonumber\\
&+\frac{3}{2}J_2\bigg(\frac{R_\star}{a_2}\bigg)^2\frac{GMm_2}{a_2}\bigg[\frac{Z_2}{\Lambda_2}-\beta_\star \sqrt{\frac{2Z_2}{\Lambda_2}}\cos(z_2)\bigg]\nonumber\\
&+\frac{Gm_1m_2}{4a_2}b_{3/2}^{(1)}\bigg(\frac{a_1}{a_2}\bigg)\bigg(\frac{a_1}{a_2}\bigg)\bigg[\frac{Z_1}{\Lambda_1}+\frac{Z_2}{\Lambda_2}\nonumber\\
&-2\sqrt{\frac{Z_1Z_2}{\Lambda_1\Lambda_2}}\cos(z_1-z_2)\bigg]\nonumber\\
&+\frac{3Gm_Gm_2}{4a_G}\bigg(\frac{a_2}{a_G}\bigg)^2\frac{Z_1}{\Lambda_1}+\frac{3Gm_Gm_2}{4a_G}\bigg(\frac{a_2}{a_G}\bigg)^2\frac{Z_2}{\Lambda_2},
\end{align}
where the first two terms represent the gravitational interaction between the planets and the host star's quadrupole. The third term represents the interaction between the inner two planets, and the final two terms arise from the distant giant's coupling to the two close-in planets. Note that we have made the assumption that $a_G\gg a_i$. 

It is convenient to introduce the following frequencies
\begin{align}\label{frequencies}
    B_{12}&\equiv n_1\frac{m_2}{4M_\star}b_{3/2}^{(1)}\bigg(\frac{a_1}{a_2}\bigg)\bigg(\frac{a_1}{a_2}\bigg)^2\nonumber\\
    B_{21}&\equiv n_2\frac{m_1}{4M_\star}b_{3/2}^{(1)}\bigg(\frac{a_1}{a_2}\bigg)\bigg(\frac{a_1}{a_2}\bigg)\nonumber\\
    \nu_{\star,i}&\equiv \frac{3}{2}n_iJ_2\bigg(\frac{R_\star}{a_i}\bigg)^2\nonumber\\
    \nu_{G,i}&\equiv n_i\frac{3m_G}{4M_\star}\bigg(\frac{a_i}{a_G}\bigg)^3.
\end{align}
These represent, respectively, the precession rate induced upon planet 1 by planet 2, the precession rate induced upon planet 2 by 1, the star-induced precession rate upon planet $i$, and the giant-induced rates.  The equations of motion may be solved in their most concise form by a canonical transformation to the complex variables \citep{morbidelli2002modern,batygin2013magnetic},
\begin{align}
    \eta_1&\equiv\sqrt{\frac{Z_1}{\Lambda_1}}\cos(z_1)+i\sqrt{\frac{Z_1}{\Lambda_1}}\sin(z_1)\nonumber\\
    \eta_2&\equiv\sqrt{\frac{Z_2}{\Lambda_2}}\cos(z_2)+i\sqrt{\frac{Z_2}{\Lambda_2}}\sin(z_2),
\end{align}
from which we may derive the inclinations
\begin{align}
    I_i=\sqrt{\frac{2Z_i}{\Lambda_i}}&=\sqrt{2\eta_i\eta_i^*}.
\end{align}
In terms of these new variables, Hamilton's equations read (see, e.g., \citealt{batygin2013magnetic})
\begin{align}
    \dot{\eta}_i=i\frac{\partial \mathcal{H}}{\partial \eta_i^{*}}\frac{1}{\Lambda_i}.
\end{align}
Given the above definitions, the dynamical evolution of the system is governed by the linear differential equation,
\begin{equation}\label{linear}
    \frac{d}{dt}\begin{pmatrix}\eta_1\\\eta_2\end{pmatrix}=iM\begin{pmatrix}\eta_1\\\eta_2\end{pmatrix}-i\frac{\beta_\star}{\sqrt{2}}\begin{pmatrix}\nu_{\star,1}\\\nu_{\star,2}\end{pmatrix},
\end{equation}
where the matrix $M$ takes the form
\begin{equation}
M = \begin{pmatrix}\nu_1+B_{12}&-B_{12}\\-B_{21} &\nu_2+B_{21}\end{pmatrix},
\end{equation}
and we define the sum of the precession rates induced by the star and giant as
\begin{align}\label{nui}
\nu_i\equiv \nu_{\star,i}+\nu_{G,i}.
\end{align}
Additional pertubers that are coplanar with the giant may be included by simply adding them to $\nu_i$, such as the disk's potential, which we describe in Section~\ref{disk}.

\subsection{Generalization to $n_p$-planets}

The dynamical equations derived in the previous section are straightforward to generalize to $n_p$ inner planets. Specifically, the matrix $M$ becomes $n_p\times n_p$ in dimensionality, with elements
\begin{align}
M_{ii}&=\nu_i+\sum_{j=1,j\neq i}^{N}B_{ij}\nonumber\\
M_{ij}&=-B_{ij}.
\end{align}\label{M}
The frequencies $B_{ij}$ are defined similarly to those in equations~\ref{frequencies},
\begin{align}
B_{ij}\equiv n_i\frac{m_j}{4M_\star} b_{3/2}^{(1)}(\alpha_{ij})\alpha_{ij}\bar{\alpha}_{ij}
\end{align}\label{M}
where $\bar{\alpha}_{jk}=1$ if $a_{j}>a_k$ and $a_j/a_k$ otherwise. Finally, we define the complex inclination vector with elements $\xi_i\approx \sqrt{2}\eta_i$, the magnitude of which being equal to the orbital inclination $I_i$, leading to the generalized 
evolution equation
\begin{align}\label{linear2}
\boxed{\frac{d \bm{\xi}}{dt}=i\bm{M}\bm{\xi}-i\beta_\star\bm{\nu}_\star}.
\end{align}
Here we have introduced the vector $\bm{\nu}_\star$ with elements $\nu_{\star,i}$ given by Equation~\ref{frequencies}. Physically, equation~\ref{linear2} describes a system of $n_p$ planetary-mass rings, perturbed by an exterior ring with zero inclination (the giant), and an inner ring of inclination $\beta_\star$ (the stellar equatorial bulge).

\section{Stellar quadrupole compared to distant giant}
\label{sec: Compare}

\subsection{Laplace surface: single inner planet case}
\label{sec: Laplace surface}

The combined perturbations from the stellar quadrupole and the distant giant cause each of the close-in planets to posses an equilibrium tilt; the steady-state configuration of orbital inclinations. In the absence of a distant giant, this equilibrium is star-aligned; whereas in the absence of a $J_2$ the equilibrium is giant-aligned. Solving Equation~\ref{linear2} for $\bm{\dot{\xi}}=0$ yields the equilibrium inclinations of each planet, given by
\begin{align}\label{steady}
\bm{\xi}_{s}=\beta_\star\mathbf{M}^{-1}\bm{\nu}_\star.
\end{align}
Note that neither the components of $\mathbf{M}$, nor $\bm{\nu}_\star$ depend upon the stellar obliquity $\beta_\star$ (Equation~\ref{frequencies}). Accordingly, within the small-angle approximation, the equilibrium inclinations scales in direct proportion to the stellar obliquity.

The simplest possible system consists of one interior planet at a semi-major axis of $a_1$. (Later in Section \ref{sec: multiple inner planets}, we will generalize to multiple interior planets.) In the single-planet case, the steady-state inclination coincides with the normal to the ``Laplace Surface" \citep{tremaine2009satellite}. Solving the one-planet case of Equation~\ref{steady} yields the equilibrium inclination 
\begin{align}\label{Laplace_Plane}
\xi_{s,1}=\frac{\beta_\star}{1+X}, 
\end{align}
where $X$ is the ratio between the star-induced and giant-induced precession rates (as defined in Equation \ref{frequencies}), 
\begin{equation}\label{X}
X\equiv \frac{\nu_{G,1}}{\nu_{\star,1}}\approx\frac{m_Ga_1^5}{2J_2 R_\star^2 a_G^3 M_\star}.
\end{equation}
The Laplace surface may be visualized as the plane about which the inner planet's orbit precesses. Specifically, if the giant dominates, $X\gg1$ and $\xi_{s,1}\rightarrow 0$. In this case, the inner planet's orbital plane remains unaltered, assuming it begins in a disk-aligned state (i.e., with $I_1$=0). On the other hand, if the star dominates, $X\ll 1$ and  $\xi_{s,1}\rightarrow \beta_\star$. In this case, the inner planets that inherit the disk's original plane will have their orbital planes altered as they precess about the star's (tilted) spin axis.

Crucially, even if the giant and star exert comparable influences, i.e., $X\sim1$, the equilibrium inclination still differs from the natal disk's plane by $\xi_{s,1}\sim \beta_\star/2$. Accordingly, the role of the stellar quadrupole is to displace the equilibrium orientations of the inner planets away from disk-aligned. In order to deduce typical values of $\xi_{s,1}/\beta_\star$, or equivalently $X$, we first estimate values of $J_2$ for young stars at the point of disk dispersal. In the next section, we compare the inferred quadrupole moments to observed examples of distant giant planets. 

\subsection{Magnitude of the stellar quadrupole}
The magnitude of $J_2$ has not yet been measured for young stars, however, the precession rates of planets orbiting the rapidly-rotating main-sequence stars WASP-33 \citep{watanabedoppler}, Kepler-13A and HAT-P-7 \citep{masuda2015spin} reveal associated values of $J_2\approx 10^{-4}$. With a much slower rotation rate, the Sun's $J_2$ is lower, at $\sim10^{-7}$ \citep{pireaux2003solar}.

 Stellar models serve as the only available estimates of $J_2$ immediately after planet formation, when rapid rotation rates drive $J_2$ to larger values. Specifically, polytopic stellar structure models may be utilized to relate $J_2$ to the stellar spin-rate and the tidal love number $k_2$ (twice the apsidal motion constant\footnote{The difference in nomenclature arose from historical application of the former to planets and the latter to stars.}), through the following approximate relationship \citep{1939MNRAS..99..451S,Ward1976secular,spalding2016spin}
\begin{align}\label{J2}
J_2&\approx \frac{1}{3}k_2\frac{\omega_\star^2}{G M_\star/R_\star^3}\nonumber\\
&\sim 10^{-3}\bigg(\frac{k_2}{0.2}\bigg)\bigg(\frac{P_\star}{\textrm{day}}\bigg)^{-2}\bigg(\frac{R_\star}{R_\odot}\bigg)^3\bigg(\frac{M_\star}{M_\odot}\bigg)^{-1},
\end{align}
where the quantity $\sqrt{G M_\star/R\star^3}$ is the break-up rotational velocity. Inserting nominal parameters for WASP-33 yields $J_2\approx 4\times 10^{-4}$ \citep{iorio2011classical}, slightly larger than that inferred empirically by \citet{watanabedoppler} but in agreement to an order of magnitude.  

\begin{figure}[!ht]
\centering
\includegraphics[trim=0cm 0cm 0cm 0cm, clip=true,width=1\columnwidth]{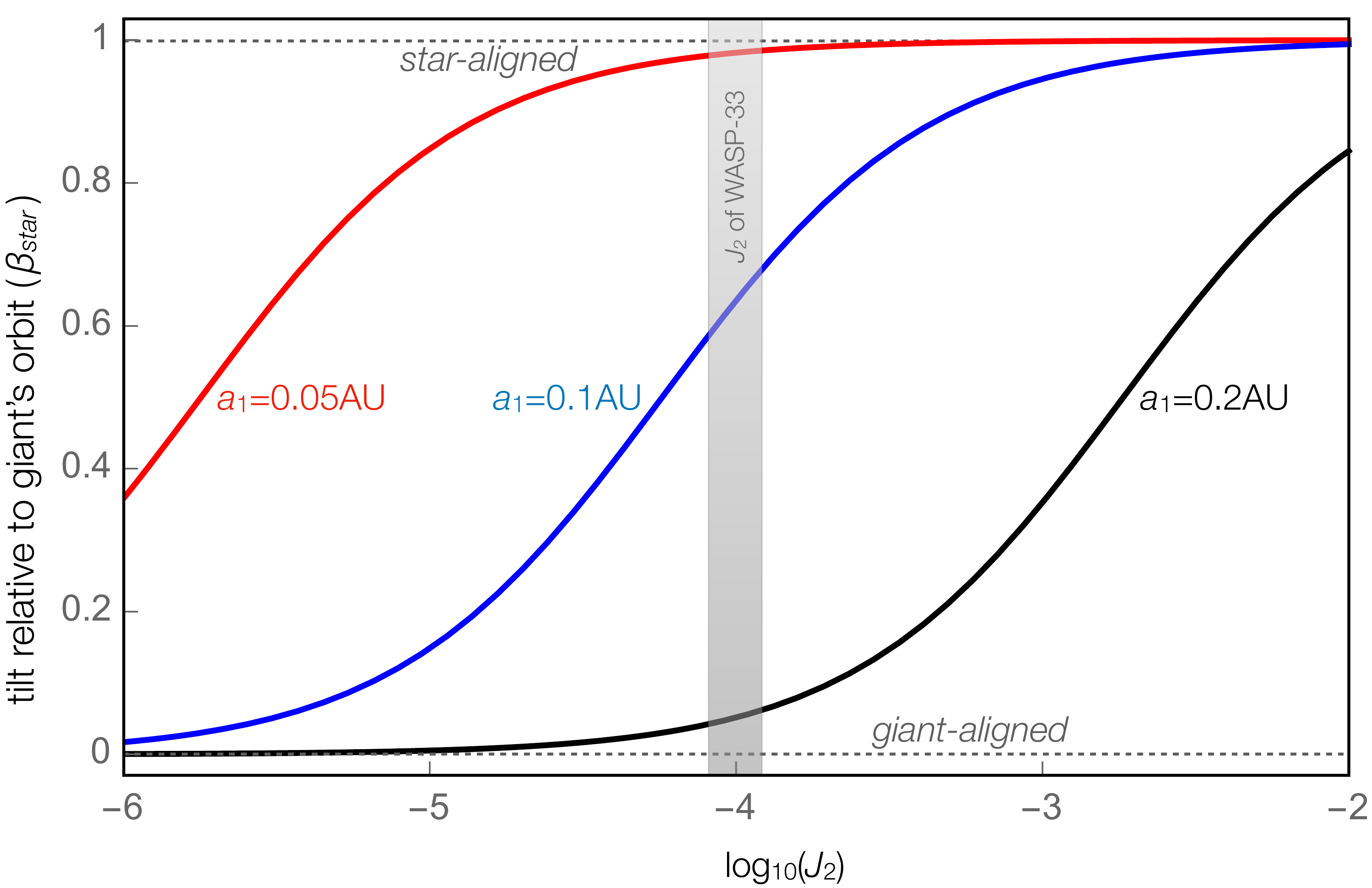}
\caption{The equilibrium inclination (Laplace plane), scaled by the stellar obliquity, experienced by a single planet orbiting interior to a giant planet and exterior to an oblate star. Three semi-major axes are denoted. The measured $J_2\approx 10^{-4}$ of WASP-33 is depicted, but values of $J_2$ during the disk-hosting stage likely far exceed this.}\label{Laplace}
\end{figure}

 Pre-main sequence stars are initially fully convective, corresponding to $k_2\approx 0.2$ \citep{batygin2013magnetic}. Eventually, those with $M_{\star}\gtrsim 0.3 \ M_\odot$ develop a radiative core, approaching $k_2\approx0.02$ in the fully radiative limit.  A Sun-like star likely remains fully convective throughout the disk-hosting stage, but more massive stars make the transition earlier on. We illustrate the equilibrium inclination of a single planet, scaled by the stellar obliquity in Figure~\ref{Laplace} for three different semi-major axes and a range of $J_2$ values. We fix the distant giant to reside on a circular orbit of $a_G=2$\,AU and possess a mass $m_G=2M_J$, which represent typical values \citep{2019AJ....157...52B,masuda2020mutual}. 
 
\begin{figure*}[!ht]
\centering
\includegraphics[trim=0cm 0cm 0cm 0cm, clip=true,width=1\textwidth]{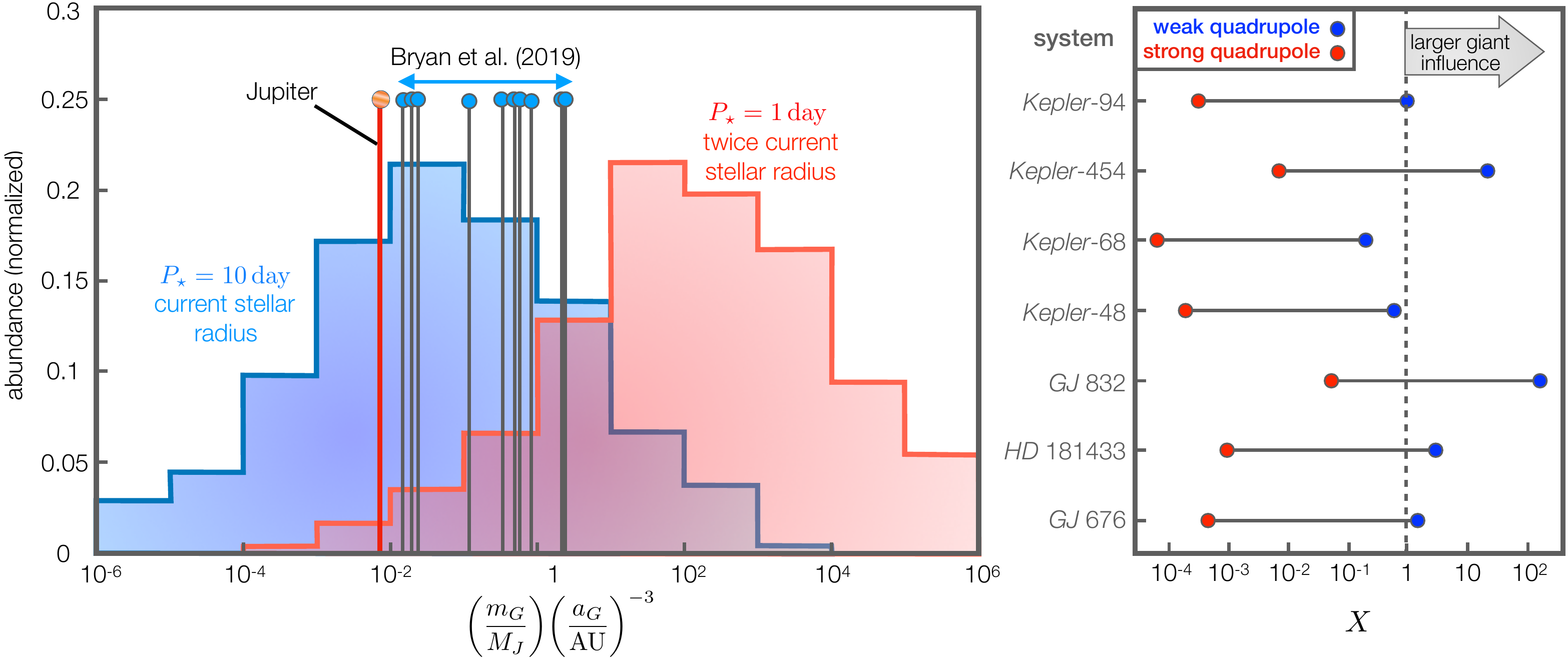}
\caption{Empirical estimates of the relative secular influences of the star and distant giant. In the left panel, we display histograms of the the tidal potential from a distant giant ($m_G/a_G^3$) required to dominate the orbital planes of observed transit singles. The blue histogram assumes the host star to have the present-day stellar radius with a 10\,day rotation period, whereas the red illustrates an expanded star with a 1\,day period. Vertical grey lines denote the values measured for 10 giants found exterior to close-in transiting planets listed in \citet{2019AJ....157...52B}, along with a red line for Jupiter. The right panel directly computes $X$ (Equation~\ref{X}) for 7 of these observed distant giants, using the orbits of their innermost planet as $a_1$. Once again, blue dots refer to a slower-rotating host star and red to the faster rotation. In general, the observed giant planets exert a comparable influence to the host star if the star rotates slower than 10\,days, but for the faster-rotation characteristic of pre-main sequence stars, the giants typically exert a sub-dominant effect.}\label{Giant_Compare}
\end{figure*}
The stellar quadrupole is comparable to the giant's influence for a wide range of parameters. Specifically, a planet with $a_p\lesssim0.05$\,AU is forced to precess approximately around the stellar spin axis, even for the relatively small $J_2\sim10^{-5}$. Planets orbiting at 0.2\,AU feel a somewhat larger influence from the distant giant, but nonetheless are driven toward roughly half of the stellar obliquity at $J_2\sim 10^{-3}$, values attainable during the first few million years of stellar evolution \citep{bouvier2014angular,spalding2016spin}.

\subsection{Relative secular influences of the star \& giant}

Having discussed the typical properties of young stars, we now compare the stellar quadrupole to the distant giant orbital parameters required to provide a comparable secular influence. Specifically, \citet{2019AJ....157...52B} indicated that up to $\sim 30\%$ of systems of close-in transiting planets are likely to possess an exterior giant with mass $0.5<m_G/M_J <20$ and orbital distance $1<a_G/\textrm{AU} <20$. Using nominal parameters of $a_G=2\,$AU and $m_G=2\,M_J$, we find that 
\begin{equation}
\label{ratio_with_numbers}
X\approx\frac{1}{20}\bigg(\frac{a_1}{0.1\textrm{AU}}\bigg)^5 \bigg(\frac{R_\star}{R_\odot}\bigg)^{-5}\bigg(\frac{m_G}{2M_J}\bigg)\bigg(\frac{a_G}{2\textrm{AU}}\bigg)^{-3}\bigg(\frac{P_{\star}}{\textrm{day}}\bigg)^2.
\end{equation}
Thus, as illustrated in Figure~\ref{Laplace}, the stellar quadrupole dominates at early times for fiducial parameters. However, the strong dependence of $X$ upon $(a_1/R_\star)^5$ indicates that wider orbits may be more affected by the exterior giant (as indeed is the case in our Solar System). Moreover, the tidal potential of the distant giant scales as $m_G/a_G^3$, which ranges over many orders of magnitude \citep{bryan2016statistics,masuda2020mutual}. 

In order to obtain a population-level idea of the relative influence of distant giants as compared to the stellar quadrupole, we turn to the NASA Exoplanet Archive \citep{2013PASP..125..989A}. We consider all confirmed, single-transiting planets possessing radii $R_p<4R_\oplus$ and obtain their semi-major axes and host star radii, when available.

This procedure yielded 1211 systems. For each system, we compute the value of $m_G/a_G^3$ that would be required to generate $X=1$ (Equation~\ref{X}). We consider two extreme cases for the stellar quadrupole. First we considered a ``strong quadrupole'' case, where we assumed that the young star's period was $P_\star=1\,$day and its radius was twice the present-day value. The second case is the ``weak quadrupole" regime, where we chose $P_\star=10$\,days and used the modern stellar radius. 

In the left panel of Figure~\ref{Giant_Compare}, we present histograms of the values of $m_G/a_G^3$ required to enforce $X=1$ in the weak (blue) and strong (red) quadrupole cases. Superimposed, we illustrate the values of 10 giant planets, with well-constrained orbits, known to reside exterior to close-in transiting planets, as listed in Table 3 of \cite{2019AJ....157...52B}. We included a red vertical line denoting Jupiter. Note that Jupiter's tidal potential is weaker than all 10 from the \cite{2019AJ....157...52B} sample. This is likely a result of biases, intrinsic to RV surveys, which favor giants that are closer-in and more massive. Accordingly, the typical influence of distant giants is likely lower than that inferred from the current observed sample.

From inspection, the tidal potentials of the observed distant giants roughly coincide with the peak of the distribution corresponding to a weak stellar quadrupole (blue), but they fall far below the histogram illustrating a strong stellar quadrupole (red). These observations suggest that the secular influence of exterior giants is at best comparable to, but is often weaker than, the stellar quadrupolar influence. 

Looking more specifically at the systems outlined by \citet{2019AJ....157...52B}, we considered 7 distant giant planets that possess inner super-Earths with well-characterized orbits\footnote{Of the 10 giants plotted in the left panel of Figure~\ref{Giant_Compare},we excluded HD 181433d as it was 1 of 2 giants in the same system, WASP-47c because it constitutes a special case of possessing a close-in hot Jupiter \citep{becker2015wasp}, and 55 Cnc d because its innermost members extended to orbital radii $<0.5\,$AU.}. Of these close-in super-Earths, we considered the innermost member and computed $X$ using the measured value of $a_1$, once again performing the calculation separately for a weak and strong stellar quadrupole. The right panel of Figure~\ref{Giant_Compare} illustrates the computed $X$ values. As with the population-level histograms, $X$ is typically smaller than unity unless the weaker limit for the stellar quadrupole is adopted, though possible exceptions are GJ 832 and Kepler-454.

In general, the observed distribution of distant giant parameters, coupled with the semi-major axes of close-in single-transiting planets, suggests that the stellar quadrupole exerts at least a comparable, but at most dominant, secular influence upon close-in planetary systems. As mentioned above, this strong quadrupolar influence tends to force inner planetary orbital planes to precess about an axis that is displaced from disk-aligned by a magnitude comparable to the stellar obliquity $\beta_\star$. Close-in planets precess faster, leading to the excitation of mutual inclinations.

\subsection{Multiple inner planets}
\label{sec: multiple inner planets}

\begin{figure*}[!ht]
\centering
\includegraphics[trim=0cm 0cm 0cm 0cm, clip=true,width=1\textwidth]{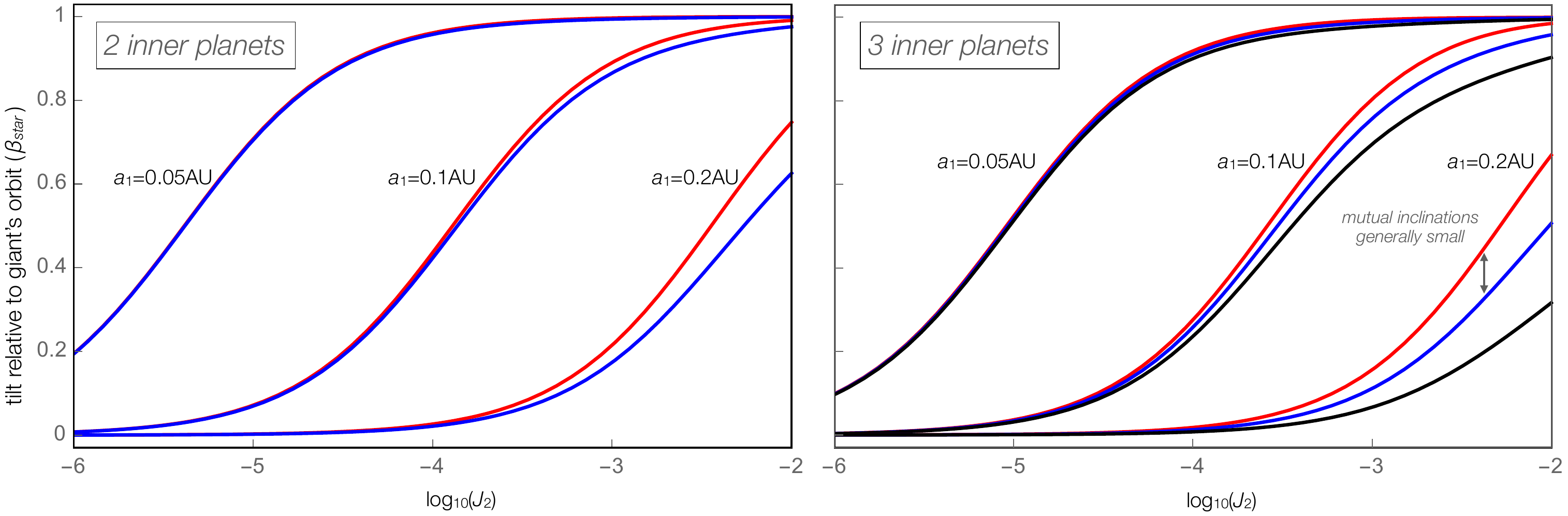}
\caption{The steady-state inclinations of systems of $n_p=2$ (left) and $n_p=3$ (right) inner planets with mass $5M_\oplus$, separated by 20 mutual Hill radii. In each case, the innermost planet orbits are set at $a_1=\{0.05,\,0.1,\,0.2\}\,$AU. In general, the equilibrium inclinations are similar to those experienced for a single planet (Figure~\ref{Laplace}). }\label{Multis}
\end{figure*}

We now generalize the steady-state treatment from Section \ref{sec: Laplace surface} to multi-planet systems by solving Equation~\ref{steady} for $n_p=2$ and $3$. As in Figure~\ref{Laplace}, we choose 3 different values for $a_1=\{0.05,\,0.1,\,0.2\}$\,AU. For simplicity, we assume the $n_p$ close-in orbits are uniformly spaced with a distance of 20 mutual Hill radii, where we define a mutual Hill radius as
\begin{align}
R_{H,mutual}\equiv\bigg(\frac{m_1+m_2}{3M_\star}\bigg)^{1/3}\frac{a_1+a_2}{2}.
\end{align} 

This spacing is typical of \textit{Kepler} compact multi-planet systems \citep{weiss2018california}. These planets also exhibit a high level of intra-system uniformity in planetary radii \citep{weiss2018california} and masses \citep{millholland2017kepler}; thus, we set all of the inner masses $m_i$ equal.  By assuming a constant separation in Hill radii, our calculations are relatively insensitive to the precise planetary mass; choosing a higher $m_i$ would simply increase the separation between adjacent planets. Accordingly, we simply choose $m_i=5\,M_\oplus$ throughout. 

We illustrate the steady-state inclinations of $n_p=2$ and $3$ planet systems in Figure~\ref{Multis}. By visual comparison to the case of $n_p=1$ in Figure~\ref{Laplace}, the equilibrium inclinations of all $n_p$ planets are similar to one-another. Accordingly, in general, all planets within an inner system with $a_1\lesssim 0.2$\,AU are initially forced to precess around a plane that differs significantly from that of the distant giant.

\subsection{Stellar spin-down}

Over time, the stellar quadrupole decays as a result of contraction, spin-down, and the formation of a radiative core. Indeed, the Sun possesses a value of $J_2\sim 10^{-7}$ \citep{park2017precession}, which is orders of magnitude lower than expected for pre-main sequence stars. Figures~\ref{Laplace} and~\ref{Multis} indicate that such small $J_2$ leads the inner planets to precess around the distant giant's plane at late times, regardless of the initial stellar $J_2$. Below, we show that stellar spin-down occurs over long-enough timescales that if the planets began star-aligned, they adiabatically evolve to giant-aligned in response to the loss of $J_2$.

\section{Dependence upon post-disk conditions}
\label{sec: Dependence}

\begin{figure*}[!ht]
\centering
\includegraphics[trim=0cm 0cm 0cm 0cm, clip=true,width=0.9\textwidth]{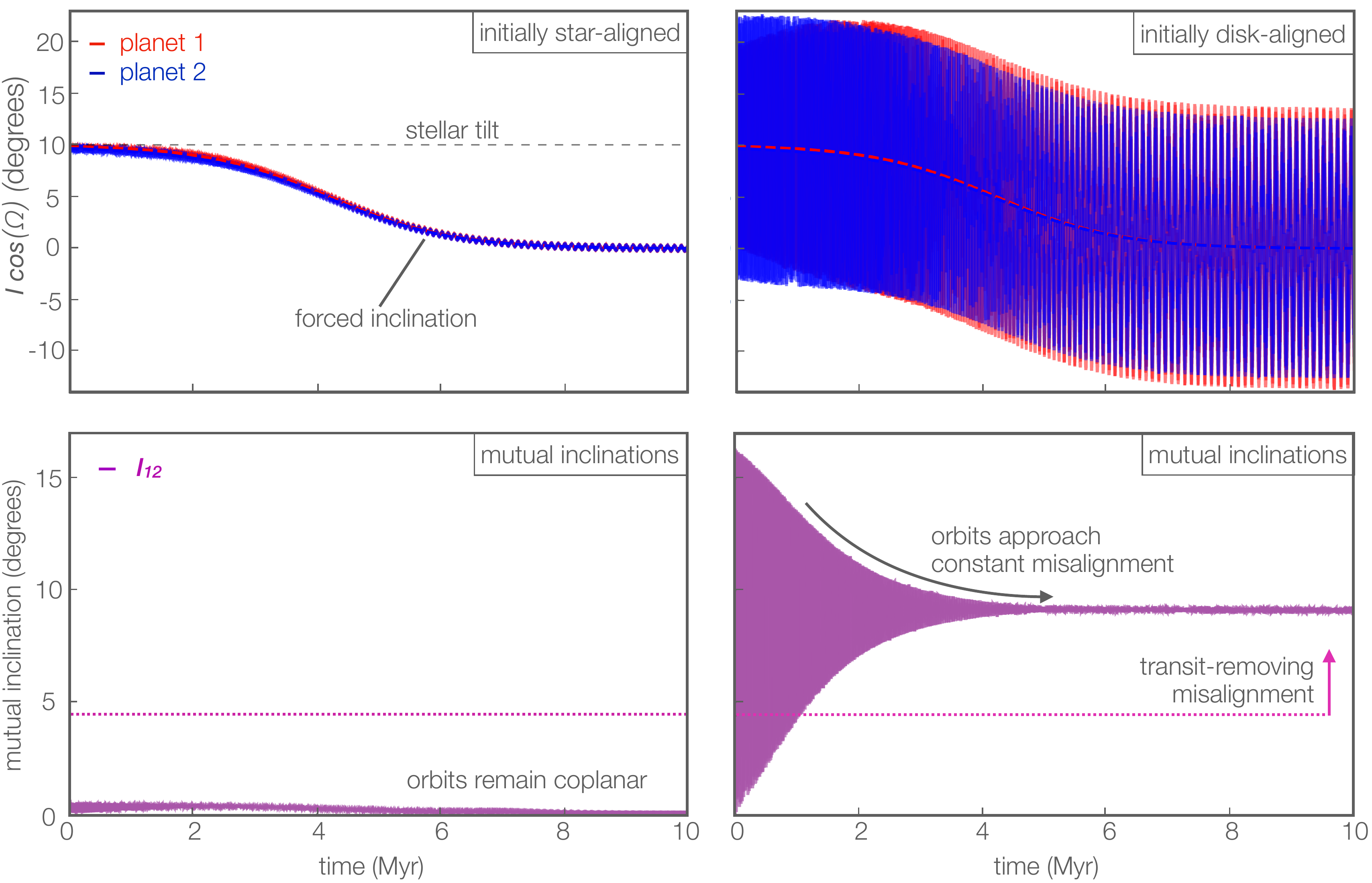}
\caption{Time evolution of the orbital inclinations (top) and planetary mutual inclinations (bottom) as the stellar $J_2$ decays for $n_p=2$ inner planets with $a_1=0.1$\,AU. The left and right panels illustrate scenarios where the inner planets are initially star-aligned and disk-aligned, respectively. The innermost planet is denoted in red and second planet in blue with mutual inclinations illustrated in purple. The horizontal dashed lines indicates the minimum mutual inclination to reduce the transit number in the lower panels, and the stellar obliquity in the top panels. Only systems where the inner planets are initially inclined with respect to the stellar equator attain and maintain mutual inclinations throughout stellar spin-down.}\label{2p}
\end{figure*}

\begin{figure*}[!ht]
\centering
\includegraphics[trim=0cm 0cm 0cm 0cm, clip=true,width=0.9\textwidth]{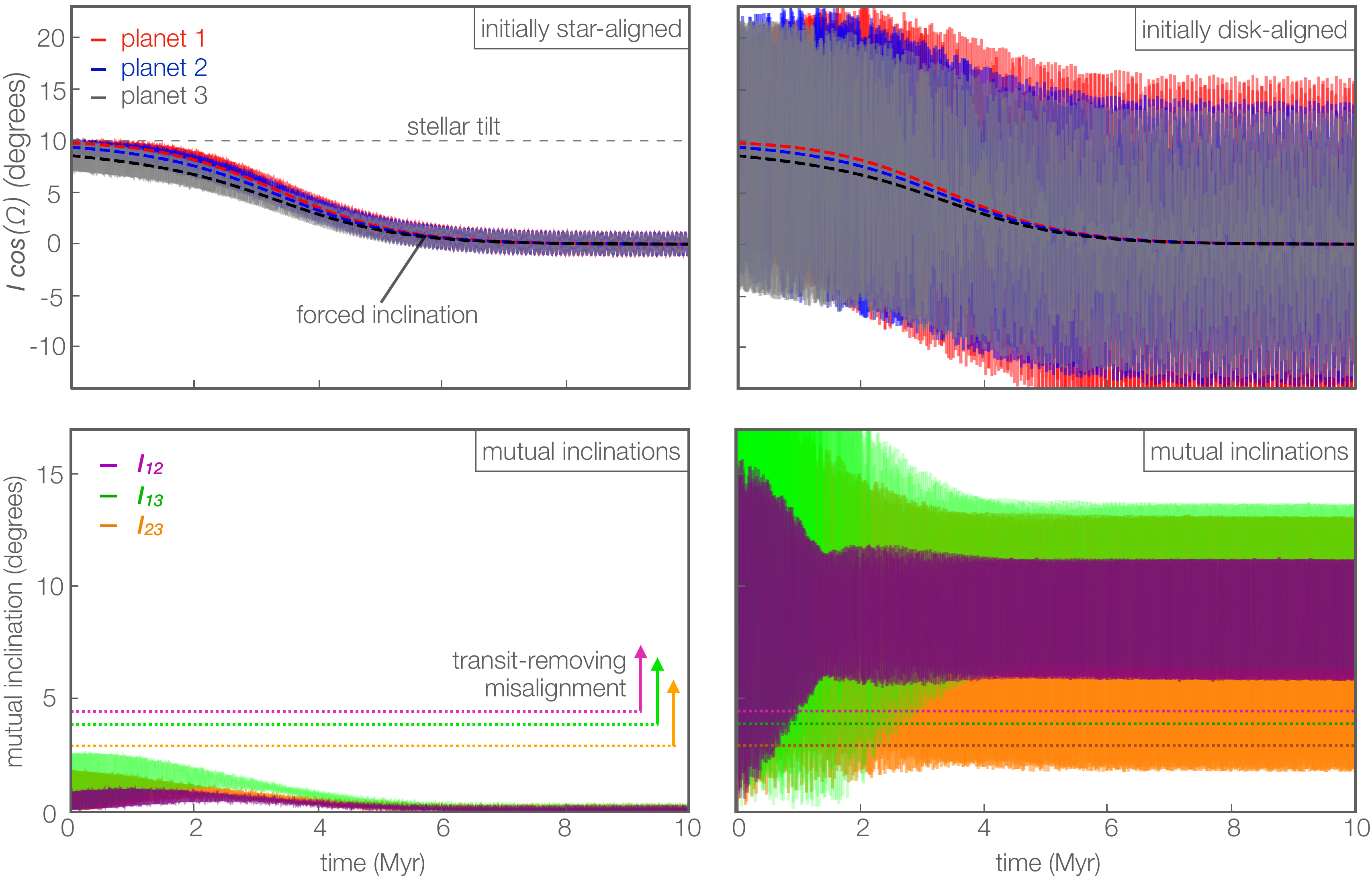}
\caption{Time evolution of the orbital inclinations (top) and planetary mutual inclinations (bottom) as the stellar $J_2$ decays for $n_p=3$ inner planets with $a_1=0.1$\,AU. As with Figure~\ref{2p}, the left and right panels illustrate initially star-aligned and disk-aligned configurations, respectively. Planets 1, 2 and 3 are assigned red, blue and gray lines in the upper panels, with dashed indicating equilibrium inclinations and solid representing the time evolution of inclinations. In the bottom panels, the mutual inclinations between planets 1-2 (Purple), 2-3 (Orange) and 1-3 (Magenta) are depicted. The horizontal dashed lines indicate the mutual inclinations that generally reduce the transit number in the lower panels, and the stellar obliquity in the top panels. }\label{3p}
\end{figure*}

In the previous section, we showed that the stellar quadrupole typically exerts a dominant secular influence upon close-in planets at early times, driving their orbits away from the plane of a distant giant. Thus if the inner system of $n_p$ planets are aligned with the stellar equator immediately following disk-dispersal (``star-aligned''), they will likely not develop large mutual inclinations with respect to one-another. On the other hand, if the inner planets retain their primordial alignment with the natal disk (``disk-aligned''), but the star possesses an obliquity, then the inner planets acquire mutual inclinations as they differentially precess about their equilibrium inclinations, which are approximately star-aligned at early times. 

In this section, we solve the fully time-dependent evolution of the secular system for these two post-disk orbital configurations: star alignment and disk alignment. To model stellar spin-down, we allow the stellar quadrupole to decay from an initial value of $J_{2,0}$, over a timescale $\tau_\star$:
\begin{align}
J_{2}(t)=J_{2,0}\exp\bigg(-\frac{t}{\tau_\star}\bigg).
\end{align}
The goal of this section is to distinguish between the two alternative conditions of star-aligned and disk-aligned, i.e., we will show that no matter how strong the stellar quadrupole is, mutual inclinations will not result unless the inner planets exhibit a primordial misalignment with respect to the stellar spin axis. Accordingly, we set $J_{2,0}=10^{-2}$, which is on the upper-end of the range of likely values. 

For the sake of simplicity, we fix $R_\star =R_\odot$, which is likely smaller than typical T-Tauri stellar radii which decrease with time \citep{gregory2016influence}. Thus, the time-dependence of the stellar quadrupole moment is contained entirely within our expression for $J_2$. Given that the evolution is adiabatic, the precise trajectory of $J_2$ over time is not important.

The mutual inclination $I_{ij}$ between planets $i$ and $j$ is considered large enough to remove the two planets from a co-transiting configuration if \citep{spalding2016spin}, 
 \begin{align}\label{Iij}
 I_{ij}\gtrsim \frac{R_\star}{a_i}+\frac{R_\star}{a_j}.
 \end{align}
This condition is a simplification; co-transits may occur at mutual inclinations above or below this limit, depending upon viewing geometry \citep{ragozzine2010value,steffen2012kepler}. Nevertheless, it serves as a convenient metric to deduce the significance of mutual inclinations excited during our simulations. 

\subsection{Star-aligned}

We first suppose that the initial orientations of each planet are star-aligned, given in terms of the complex inclination vector as
\begin{align}
\xi_{i}\big|_{t=0}&=\beta_\star,
\end{align}
i.e., each orbit normal is parallel to the stellar spin axis. For the sake of definiteness, we set the stellar obliquity to $\beta_\star=10^{\circ}$ (similar to the Sun's $7^{\circ}$). We set $\tau_\star=1\,$Myr and integrate equations~\ref{linear} for 10\,Myr, or 10 e-folding times of the stellar quadrupole (which approximately amounts to a drop from $J_2 = 10^{-2}$ to $J_2<10^{-6}$). 

We integrate systems of $n_p=2$ and $3$ inner planets, choosing the innermost $a_1=0.1$\,AU, separating the planets by 20 mutual Hill spacings, as above. We present the time evolution of the planetary inclinations and their mutual inclinations in the left-panels of Figure~\ref{2p} ($n_p=2$) and Figure~\ref{3p} ($n_p=3$). In the upper left panels, colored solid lines depict the real part of the complex inclinations $\xi_i$, whereas colored dashed lines illustrate the equilibrium inclinations computed in the previous section. As the stellar quadrupole diminishes, the planets adiabatically reorient from star-aligned to giant-aligned. In the bottom left panels, we plot the mutual inclinations between all pairs of planets, which remain below 1$^{\circ}$. Accordingly, in the star-aligned initial state, the inner planets remain coplanar. 

\subsection{Disk-aligned}

Now suppose that the inner planets remain aligned with the disk's plane subsequent to disk-dispersal, such that
\begin{align}
\xi_{i}\big|_{t=0}&=0.
\end{align}
As before, we simulate systems with $n_p=2$ and $3$ for 10\,Myr and present the results in the right panels of Figures~\ref{2p} and~\ref{3p}. In contrast to the left panels, a disk-aligned initial condition drives significant mutual inclinations between the inner planets (bottom-right panels of Figures~\ref{2p} and~\ref{3p}). Critically, throughout the subsequent evolution, these mutual inclinations remain high, exceeding the magnitude required to reduce the observed transit number (the horizontal dashed lines; Equation~\ref{Iij}). It is important to emphasize that the distant giant did not \textit{cause} the mutual inclinations, but still ended up inclined with respect to the inner planets.

Notice that as the stellar quadrupole decays, the planetary mutual inclinations decrease somewhat. This feature implies that while the initial, large $J_2$ drives mutual inclinations, later stellar spin-down plays a partial role in dynamically-cooling close-in planetary systems. Similar behaviour was highlighted in an analogous scenario by \citet{anderson2018teetering}, whereby stellar spin-down reduced the mutual inclination between a warm Jupiter and an exterior perturbing giant companion. Accordingly, in general, the mutual inclinations observed within modern-day, close-in planetary systems were likely once larger. The same may be said of satellite systems around giant planets which, like stars, lose their quadrupole moments with time \citep{batygin2018terminal}. We defer a detailed discussion of this process to future work. 

\begin{figure*}[!ht]
\centering
\includegraphics[trim=0cm 0cm 0cm 0cm, clip=true,width=0.9\textwidth]{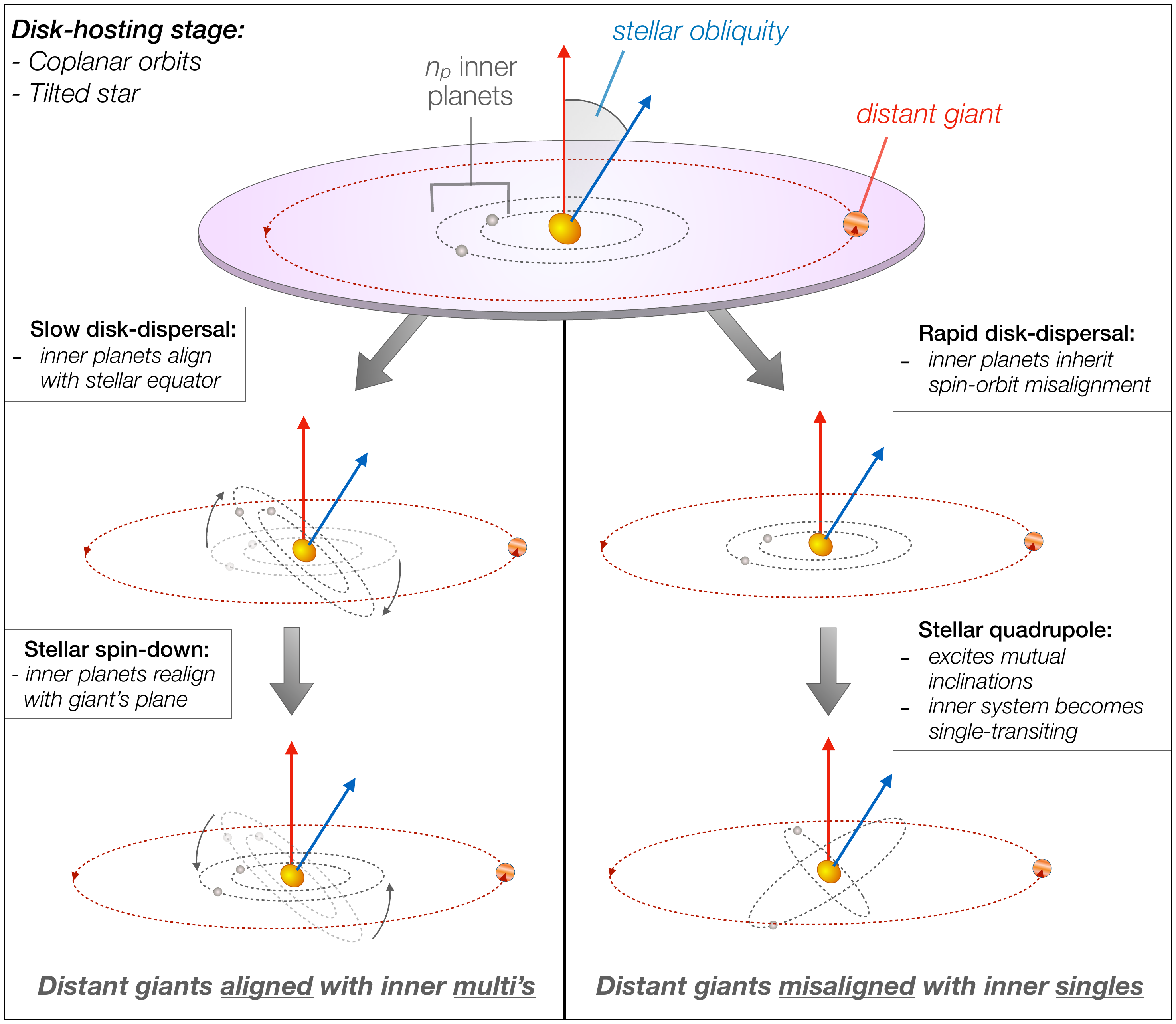}
\caption{A schematic illustration of the inclination evolution described in this work. A system of $n_p$ close-in planets orbits in the same plane as an exterior giant planet while the protoplanetary disk is present (top). We assume the star possesses a non-zero tilt relative to the disk's plane. From there, the disk either disperses rapidly (right) or slowly (left). If dispersal is rapid, the planets inherit the disk's plane and find themselves subject to the inclined quadrupolar potential of the star, at which point they are driven into misalignment with one another. If dispersal is slow (left), the inner planets have time to adiabatically reorient to star-aligned, preventing the excitation of mutual inclinations. During subsequent stellar spin-down, the stellar quadrupole is lost, and the planets reorient back to the exterior giant's orbital plane. Low inclinations between distant giants and multi's, but larger misalignments with singles, are consistent with recent observations \citep{masuda2020mutual}.}\label{schematic}
\end{figure*}

The primary findings of this section are summarized schematically in Figure~\ref{schematic}. Specifically, an initial configuration of $n_p$ planets inclined with respect to the stellar spin axis (but aligned with a distant giant) acquires substantial mutual inclinations. The converse scenario of initial alignment with the star leads to negligible mutual inclinations among close-in planets and alignment with the distant giant as $J_2$ decays. Accordingly, the picture described here is fully consistent with the inference by \citet{masuda2020mutual}--that giants orbiting exterior to single-transiting systems are inclined, whereas those exterior to multi-transiting close-in systems are well-aligned. The question is then: what determines whether star-aligned or giant-aligned initial conditions prevail subsequent to planet formation? In the next section, we will show that the disk dispersal timescale is the primary governing factor. \\

\section{Inclination excitation during disk-dispersal}
\label{sec: disk-driven initial conditions}

During the protoplanetary disk stage, planets are expected to possess a coplanar architecture over multi-AU scales \citep{2017ApJ...840...23L,2018MNRAS.477.5104C}. In contrast, numerous physical mechanisms are capable of exciting misalignments between the disk's plane and the stellar equatorial plane, including torquing from a binary companion \citep{batygin2012primordial,spalding2014early,spalding2015magnetic,zanazzi2018planet}, magnetic star-disk interactions \citep{lai2014star}, and turbulence in the stellar core \citep{bate2010chaotic,spalding2014alignment,fielding2015turbulent}. These misalignment pathways are often invoked to explain large ($\gtrsim 30^{\circ}$) spin-orbit misalignments in exoplanetary systems, although even Sun-like obliquities of around 10$^\circ$ are sufficient to drive significant mutual inclinations (Figures~\ref{2p} \&~\ref{3p}). We thus assume that near the end of the disk-hosting stage, the central star exhibits a non-zero obliquity, while the inner planets and distant giant are coplanar. 

At the end of its lifetime, the disk disperses over a finite timescale $\tau_d$ \citep{alexander2006photoevaporation,alexander2014dispersal} that is much shorter than its lifetime \citep{haisch2001disk,mamajek2009initial}. The timescale of dispersal impacts the planets' subsequent inclination evolution, which may be understood by considering the timescale extremes. 

If the disk were to disperse instantaneously, the inner planets would inherit the disk's plane, becoming inclined to the stellar equator. The planets would then spontaneously precess around their new, star-aligned equilibrium (assuming the stellar quadrupole to dominate), exciting mutual inclinations in the process\footnote{An analogous scenario has been proposed in order to generate the inclination of Iapetus' orbit by way of the rapid dispersal of Saturn's circumplanetary disk \citep{1981Icar...46...97W}.} (Figure~\ref{schematic}). On the other hand, consider a scenario where the disk-dispersal timescale is arbitrarily long. In this case, the equilibrium inclinations of the inner planets move from disk-aligned to star-aligned slowly, allowing the inner planetary orientations to simply track their instantaneous equilibria. In doing so, the inner planets remain coplanar and align with the stellar equator. These dichotomous outcomes are depicted in Figure~\ref{schematic}. 

Referring to Figures~\ref{2p} and \ref{3p}, a rapid disk dispersal generates disk-aligned initial conditions (right panels), whereas a slow dispersal leads to star-aligned initial conditions (left panels). Accordingly, the critical factor influencing the mutual inclinations is the rapidity of the disk's dispersal. In this section, we add the disk's gravitational potential to our secular model and allow it to decay over a timescale $\tau_d$. Generally, we find that $\tau_d$ must be shorter than the period of the slowest eigenvalue of the system (typically $10^{3-4}$ years) for large mutual inclinations to result.  We discuss probable disk-dispersal timescales to Section~\ref{sec: Timescale}. 

\subsection{Adding the disk potential}\label{disk}

In order to simulate inclination excitation during disk dispersal, we model the secular potential of the disk by way of a precession frequency $\nu_{d,i}$, given by \citep{hahn2003secular}
\begin{align}
\nu_{d,i}=n_i\frac{\pi \Sigma a_i^2}{M_\star\beta},
\end{align}
which is added to $\nu_i$ in Equation~\ref{nui}. We define the aspect ratio of the disk, $\beta\approx0.05$, and the surface density, $\Sigma$. This frequency is then used to compute the diagonal elements of the updated matrix $\mathbf{M}$.

 The disk's surface density as a function of distance and time, $\Sigma(a,t)$, is chosen to follow an infinite Mestel disk \citep{mestel1963galactic,binney2011galactic,schulz2012gravitational}:
\begin{align}
\Sigma(a,t)=\Sigma_0(t)\bigg(\frac{a_0}{a}\bigg).
\end{align}
The scaling factor $\Sigma_0(t)$ is allowed to decay exponentially over a timescale $\tau_d$ following
\begin{align}\label{Mestel}
\Sigma_0(t)=\Sigma_{0,0}\exp\bigg(-\frac{t}{\tau_d}\bigg).
\end{align}
We choose $a_0=0.2\,$AU and $\Sigma_{0,0}=1300$\,g\,cm$^{-2}$ approximately in-keeping with the minimum-mass extrasolar nebula \citep{chiang2013minimum}. This is sufficient to enforce initial coplanarity with the disk plane.

\subsection{Sensitivity to disk-dispersal timescale}

\begin{figure}
\centering
\includegraphics[trim=0cm 0cm 0cm 0cm, clip=true,width=1\columnwidth]{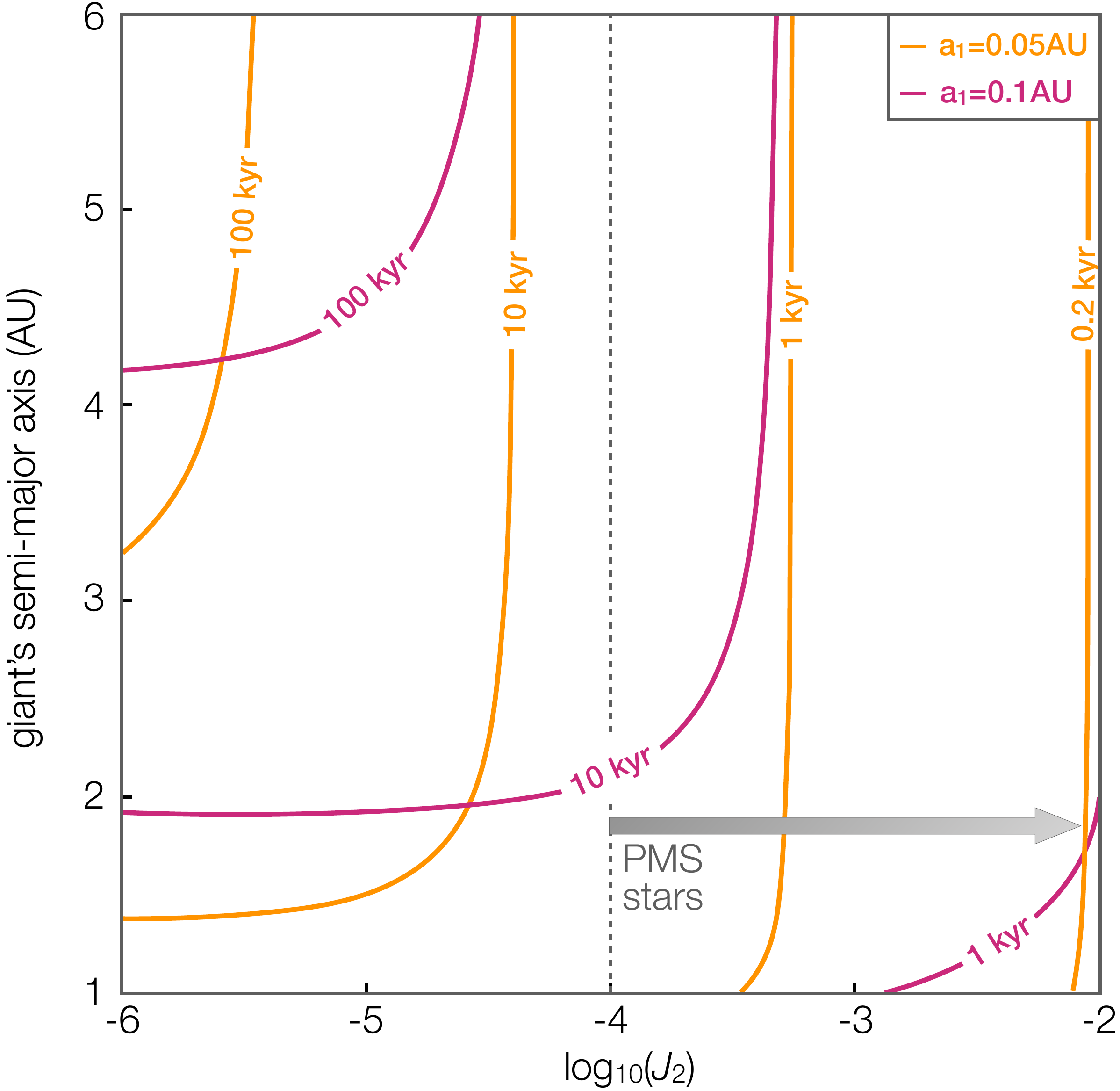}
\caption{The timescale of the slowest eigenmode as a function of the stellar quadrupole and exterior giant's semi-major axis. The giant's mass is fixed at 2$M_J$. Above  $\sim2$\,AU, the timescale depends mostly upon the stellar $J_2$. At values of $J_2$ between $10^{-2}$ and $10^{-4}$, typical for the end of disk-dissipation, the slowest eigenmode typically lies between 1-10\,kyr in period. Accordingly, if the disk disperses over a longer timescale than this, the planets will approximately maintain their equilibrium inclinations from the disk phase. However, a more rapid disk-dispersal leads to impulsive excitation of mutual inclinations. Note that the timescale of disk dissipation is distinct from the age of the disk (around 3\,Myr), but refers to the timescale over which the disk's quadrupolar dominance diminishes near the end of the disk's lifetime.}\label{EigenSlow}
\end{figure}

Heuristically, disk dispersal should be adiabatic if $\tau_d$ is shorter than the average precession timescale of the inner $n_p$ planets. This average rate roughly corresponds to the minimum eigenvalue $\lambda_{\mathrm{min}}$ of matrix $\mathbf{M}$. In Figure~\ref{EigenSlow}, we illustrate typical values of $\lambda_{\mathrm{min}}^{-1}$ using systems of $n_p=3$ inner planets with $a_1=0.05$\,AU and $0.1$\,AU. As can be seen, the timescale $\lambda_{\mathrm{min}}^{-1}$ can be as long as $\sim10^5$\,years or as short as $\sim10^3$\, years, depending upon the giant's semi-major axis and the stellar $J_2$. Given the large $J_2\gtrsim10^{-3}$ expected of young stars, we expect the system to evolve adiabatically if the disk's mass in the inner regions is removed over a timescale of $\tau_d\gtrsim 10^{3-4}$ years. This is roughly in-keeping with the viscous timescale of the inner 0.2\,AU of the disk gas, which we return to in the discussion (see Equation~\ref{viscous}). 

\begin{figure*}[!ht]
\centering
\includegraphics[trim=0cm 0cm 0cm 0cm, clip=true,width=1\textwidth]{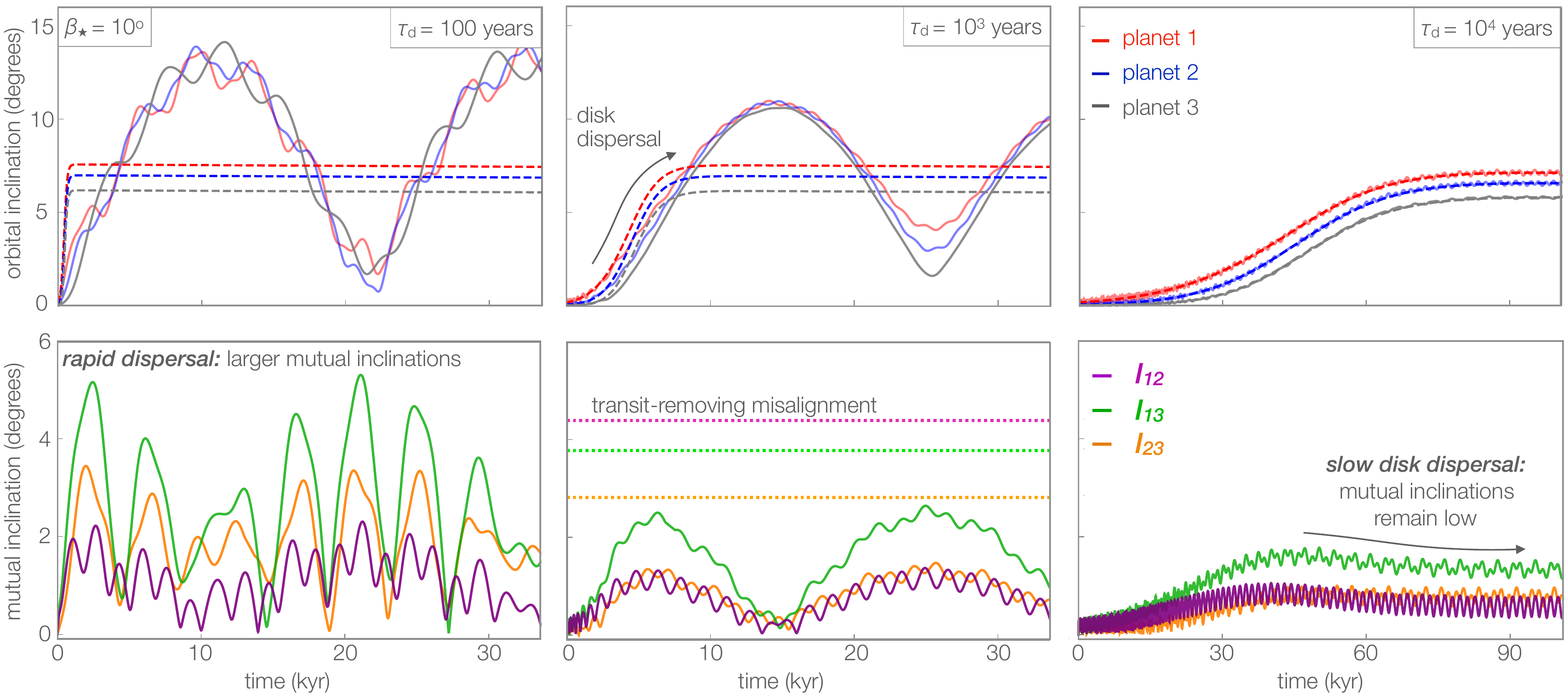}
\caption{The evolution of orbital inclinations (top) and mutual planet-planet inclinations (bottom) as the disk's mass disperses over 3 different timescales; from left to right $\tau_d=\{10^2,\,10^3,\,10^4\}\,$years. The dotted lines in the upper panels denote the equilibrium inclinations, which move from $0^{\circ}$ at large disk masses to nearly star-aligned at stellar disk masses (the stellar obliquity here is $10^{\circ}$). The lowest eigenvalue is associated with a timescale of $\sim3000$\,years. Only when disk dissipation timescales are much shorter than $\lambda_{\mathrm{min}}^{-1}$ are mutual inclinations excited that approach the stellar obliquity. For example $\tau_d=100$\,years can drive mutual inclinations up to potentially $5^{\circ}$, which is half of the stellar obliquity. The mutual inclinations required to remove planet pairs from a co-transiting configuration are marked by horizontal dashed lines in the bottom middle panel. For rapid disk dispersal timescales, $\beta_\star=10^\circ$ is marginally sufficient to reduce the transit number.}\label{DiskDiss}
\end{figure*}

We test the dependence of planet-star inclinations upon $\tau_d$ by performing a set of secular integrations that include the disk's secular potential. Specifically, we choose a system with $a_1=0.1\,$AU, $n_p=3$ inner planets and an initial $J_{2,0}=10^{-3}$ (indicative of $P_\star=1\,$day; Equation~\ref{J2}). These parameters correspond to an initial $\lambda_{\mathrm{min}}^{-1}\approx 3400$\,year. We simulate three different disk-dispersal timescales; a rapid case where $\tau_d=100$\,years, a marginally-adiabatic case with $\tau_d=1000$\,years, and a fully-adiabatic case where $\tau_d=10^4$\,years. The stellar obliquity is fixed at $\beta_\star=10^{\circ}$ for each run, but note that in the Laplace-Lagrange approximation used here, all angles scale roughly linearly with $\beta_\star$.  

The evolution of the 3 inner planetary orbits during disk dispersal are presented in Figure~\ref{DiskDiss}. Orbital inclinations are presented in the top panels, with dashed lines tracking the equilibrium inclinations (Equation~\ref{steady}). As the disk disperses, the equilibrium inclinations transition from disk-aligned to a value of around 8$^\circ$, close to the stellar obliquity of $10^\circ$. Mutual inclinations between all inner planet pairs are plotted in the lower panels. The adiabatic case of a slowly-dispersing disk is presented in the right-most panel. Here, all 3 inner planets closely-follow their forced equilibria, maintaining coplanarity. Thus, slow disk-dispersal yields coplanar systems that are aligned with the distant giant. 

For shorter disk dispersal timescales (middle and left panels) the inner planets acquire their inclinations more impulsively. When the disk's dispersal is marginally adiabatic ($\tau_d=1000$\,years; middle panel) the planets track their equilibrium inclinations less exactly as compared to $\tau_d=10^4$ years, but they nevertheless maintain small mutual inclinations with one another. Only planets 1 and 3 develop mutual inclinations exceeding $\sim 2^\circ$; the typical rms spread of mutual inclinations in \textit{Kepler} systems \citep{fabrycky2014architecture}. Dispersal must be strongly non-adiabatic in order to generate mutual inclinations exceeding half of the stellar obliquity of 10$^\circ$, as illustrated by the case with $\tau_d=100$\,years.

 \subsection{Reduction of transit number}
  
\begin{figure*}[!ht]
\centering
\includegraphics[trim=0cm 0cm 0cm 0cm, clip=true,width=0.85\textwidth]{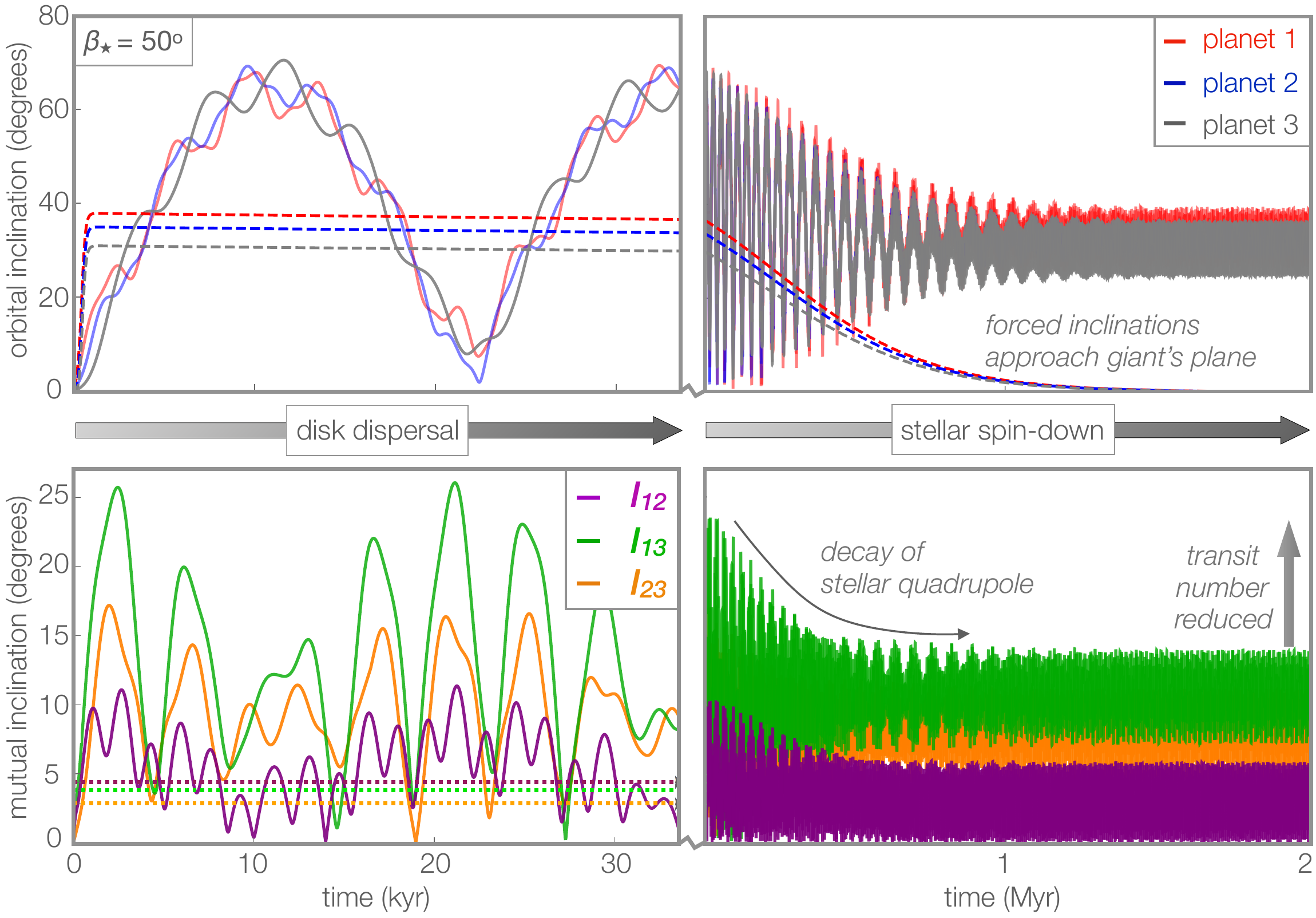}
\caption{The evolution of orbital inclinations (top) and mutual planet-planet inclinations (bottom) as the disk's mass disperses over $\tau_d=10^2\,$years. Similarly to Figure~\ref{DiskDiss}, the dotted lines in the upper panels denote the equilibrium inclinations. Here we set the stellar obliquity to $\beta_\star=50^\circ$. In order to show the evolution during and after disk dispersal, the left half of the plot is illustrated over a smaller timescale than the right half. As the stellar $J_2$ decays (beginning at $10^{-3}$), the mutual inclinations excited during disk dispersal are reduced somewhat, such that planet pairs $2-3$ and $1-3$ are removed from a mutually-transiting configuration, whereas planets $1-2$ may or may not co-transit.}\label{30deg}
\end{figure*}

When the stellar obliquity is set to $\beta_\star=10^\circ$, the most rapid disk-dispersal timescale of 100\,years generates mutual inclinations that are only barely large enough to exceed the co-transiting criterion (horizontal, dotted lines in Figure~\ref{DiskDiss}). We repeat the secular simulation above, this time with a larger stellar obliquity of $50^\circ$. The results are presented in Figure~\ref{30deg}, where we illustrate the rapid disk dispersal time of $\tau_d=100$\,years. Moreover, we continue the simulation throughout the subsequent decay of the stellar quadrupole (with $\tau_\star=0.25\,$Myr in this case). At the larger tilt of $\beta_\star=50^\circ$, mutual inclinations more often exceed the coplanarity criterion (alternatively, a larger $J_{2,0}=10^{-2}$ coupled with a smaller stellar obliquity suffices for generating non-transiting configurations; Figure~\ref{2p}).
 
In general, our secular calculations show that stellar obliquities exceeding $\beta_\star \sim 10^\circ$, coupled with rapid disk-dispersal are able to reduce the transit multiplicities of planetary systems. Coplanarity is retained if the disk disperses slowly. Subsequently, the planets are finally drawn back to their original plane--aligned with the giant--as the stellar quadrupole decays. 

Though the secular approach is critical for developing insight, it suffers several drawbacks. First, Laplace-Lagrange secular theory loses accuracy at high inclinations \citep{murray1999solar}, such that our simulations with $\beta_\star =50^\circ$ are unreliable at a quantitative level. Second, the secular frequencies of planetary systems vary with time as the disk disperses and $J_2$ changes \citep{nagasawa2005dynamical}. This process can raise eccentricities and inclinations, or even drive instabilities \citep{Ward1981solar,spalding2018resilience}. Lastly, we assumed that the giant's orbit and stellar spin axes are fixed in time, which is not correct in detail. Given these shortcomings, in the next section we check the validity of our analysis by turning to \textit{N}-body simulations.
 
\section{\textit{N}-body simulations}
\label{sec: Nbody}

Here we perform full \textit{N}-body simulations that capture the dynamical evolution of a system of close-in planets and a distant giant, beginning with disk dispersal and continuing throughout stellar spin-down. Unlike the secular approaches adopted thus far, we will allow the central star's spin-axis and the giant to freely evolve.

\begin{figure*}[!ht]
\centering
\includegraphics[trim=0cm 0cm 0cm 0cm, clip=true,width=0.9\textwidth]{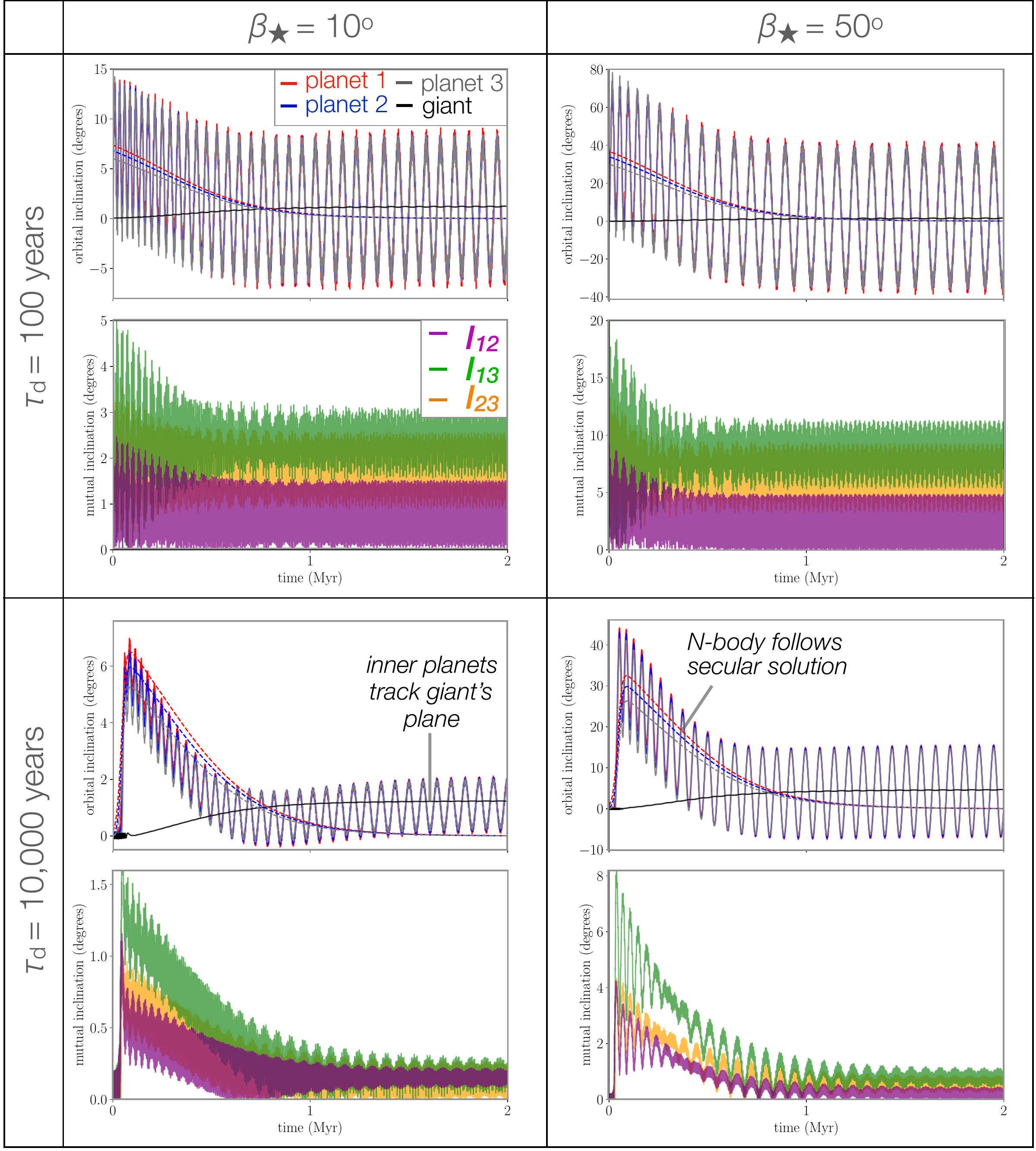}
\caption{Results of the $N$-body simulations with four different configurations. The left and right columns correspond to the configurations with the initial stellar obliquity set at $\beta_\star = 10^{\circ}$ and $50^{\circ}$, respectively. The top and bottom rows correspond to the fast and slow disk dissipation timescales, $\tau_d = 100$ years and 10,000 years, respectively. Within each configuration (quadrant of this figure), we show the evolution of orbital inclinations (top) and mutual planet-planet inclinations (bottom) as the disk's mass decreases and as the stellar quadrupole decays. Similar to Figures~\ref{DiskDiss} and \ref{30deg}, the dotted lines denote the equilibrium inclinations according to the secular solutions. The $N$-body simulations agree well with the secular solutions, even for higher stellar obliquities (compare the right column with Figure~\ref{30deg}). The only substantial difference arising within the $N$-body case is that the giant's orbit tilts slightly in response to the stellar spin-down. With a slowly-dispersing disk, the planets approach the giant's plane, as opposed to the zero-inclination state. \\
\\
\\
\\
}\label{N-body_sim_results}
\end{figure*}

Our direct numerical integrations use instantaneous accelerations in the framework of \cite{2002ApJ...573..829M}, and the planetary orbits are evolved in hierarchical (Jacobi) coordinates. Further details of the code may be found in the Methods section of \cite{2019NatAs...3..424M}.  In addition to the standard Newtonian gravitational accelerations, the bodies also experience accelerations due to the quadrupolar gravitational potential of the star and the gravitational influence of the protoplanetary disk.

The acceleration on planet $i$ due to the stellar quadrupole is given by
\begin{flalign}
&\bm{a}_{Q,i} = \frac{k_{2}}{2}\frac{{R_{\star}}^5}{r^4}\left(1+\frac{m_i}{M_{\star}}\right) \\
&\times\left[\left(5(\bm{\omega_{\star}}\cdot\bm{\hat{r}})^2 - \lvert\bm{\omega_{\star}}\rvert^2 -  12\frac{G m_i}{r^3} \right)\bm{\hat{r}} - 2(\bm{\omega_{\star}}\cdot\bm{\hat{r}})\bm{\omega_{\star}}\right],&& \nonumber
\end{flalign}
where $\bm{\omega}_{\star}$ is the star's spin vector and $\bm{r}$ is the relative position vector from the star to planet $i$. We do not account for accelerations due to the planets' own quadrupolar moments, as they are negligible. We track the evolution of the stellar spin vector, given by
\begin{equation}
I_{\star}\bm{\dot{\omega}_{\star}} = \sum_{i=1}^{n_{p}+1} -\frac{M_{\star}m_i}{M_{\star} + m_i}\bm{r} \times \bm{a}_{Q,i}, 
\end{equation}
where $I_{\star}$ is the fully dimensional moment of inertia, and the summation to $n_{p}+1$ is the total number of planets, including the distant giant.

As in the secular integrations, we model the protoplanetary disk using a Mestel disk profile (Equation~\ref{Mestel}). We assume the disk to be infinitely flat and wide (i.e. outer radius $\rightarrow \infty$), with a gravitational potential given by \citep{thommes2008dynamical,schulz2012gravitational}
\begin{align}
\phi_d=2\pi G \Sigma_0 (t)a_0 \ln\big(|z|+\sqrt{z^2+R^2}\big).
\end{align}
The disk generates accelerations on the planets in the radial and vertical directions equal to
\begin{align}
a_R&=-\frac{\pi^2G \Sigma_0(t)a_0}{R}\bigg(1-\frac{|z|}{\sqrt{R^2+z^2}}\bigg)\nonumber\\
a_z&=-\pi^2G \Sigma_0(t)a_0\frac{\textrm{sgn}(z)}{\sqrt{R^2+z^2}}.
\end{align}

Having specified the accelerations within the direct integrations, we now define the set-up and initial conditions. As with the secular integrations, we adopt a system of $3$ inner planets, each of mass $5 \ M_{\oplus}$, and set $a_1=0.1$\,AU with orbital separations of $20$ mutual Hill radii. The distant giant is set at $a_G=2$\,AU with a mass $m_G=2 \ M_J$. All planets are initialized with small inclinations of $0.1^{\circ}$ and randomized orbital orientations. We initialize $P_{\star}=1\,\textrm{day}$, $R_\star=1 \ R_\odot$, and $k_2=0.2$, leading to an initial $J_{2,0}\approx 1\times 10^{-3}$. 

As before, $J_2$ is allowed to decay over a timescale of $\tau_\star=0.25\,$Myr. We set the initial disk surface density $\Sigma_{0,0}=1300$\,g\,cm$^{-2}$ at  $0.2\,$AU \citep{chiang2013minimum}, decaying over a timescale of either $\tau_{d}=10^2\,$years or $10^4\,$years (see Equation~\ref{Mestel}). We consider two different initial stellar obliquities, $\beta_{\star} = 10^{\circ}$ and $50^{\circ}$. Thus, we present four simulations in total, one for each combination of the two stellar obliquities and disk decay timescales. These simulations are not intended to be exhaustive, but rather to serve as a test of the secular arguments made above.

The results of our $N$-body simulations are presented in Figure \ref{N-body_sim_results}. Each quadrant contains one of the four simulation configurations. Overall, we observe strong agreement between the secular and $N$-body integrations, supporting the accuracy of our earlier results. The $N$-body simulations, however, allow us to account for two limitations of the secular model. First, the secular model assumed the giant's orbit was fixed, but here we observe that its inclination changes due to the mutual precession between its orbit and the stellar spin. Accordingly, in the slow-dispersing case, with  $\tau_d = 10^4$ years, the inner planets realign with the true plane of the giant, rather than the zero plane. 

Second, the secular code assumed small inclinations, while the $N$-body code allows for arbitrary values. The $\beta_{\star} = 10^{\circ}$ case agrees almost exactly with the secular results (Figure~\ref{DiskDiss}). Moreover, we observe that nothing qualitative changes in the high obliquity case (compare to Figure~\ref{30deg}). The final mutual inclinations in the slowly-dispersing, $\beta_{\star} = 50^{\circ}$ case are still consistent with a mutually-transiting configuration. While previous work suggests that orbital instabilities may be triggered at obliquities exceeding $\beta_\star \gtrsim40^\circ$ \citep{spalding2018resilience}, we do not observe any such instabilities here. This may be due to the anchoring influence of the distant giant planet, a topic that we will reserve for future work.

\section{Discussion}
\label{sec: discussion}

In this work, we have compared the influence of the stellar quadrupolar potential upon close-in planetary systems to the secular influence of an exterior giant planet. We found that soon after disk-dispersal, the host star typically exerts a stronger potential due to its rapid rotation, and is therefore better able to generate mutual inclinations among close-in planetary orbits (see Figure~\ref{Giant_Compare}). Throughout our analysis, we have assumed that, while the stellar spin axis is initially tilted from the disk's plane, the exterior giant's orbit remains coplanar with the disk immediately subsequent to disk-dispersal. Accordingly, even if the giant's secular potential dominated early-on, it would not generate mutual inclinations among the inner planets \citep{2017AJ....153...42L}.

If, on the other hand, the giant obtains a large inclination at a later epoch (and after stellar spin-down), through mechanisms such as planet-planet scattering \citep{chatterjee2008dynamical}, this could generate mutual inclinations among the inner planets. Such a dynamically-impulsive origin is likely the case for several specific systems, such as $\pi$ Men \citep{2020arXiv200701871X} and HAT-P-11 \citep{yee2018hat}, which host exterior giants that exhibit large eccentricities and inclinations, hallmarks of dynamical instabilities.

Despite the potential for such dynamical interactions later-on, we focused upon the epoch immediately following disk-dispersal, when primordial giant-disk alignment is expected. However, in our modelling above, we simply assumed that the star possessed an initial obliquity with respect to the disk's plane. Thus, it is important to discuss more fully the likelihood that such primordial obliquities exist.

\subsection{Primordial stellar obliquities}

Stellar obliquities spanning the full range from $0-180^\circ$ have been detected in a diverse collection of planetary systems \citep{winn2010hot,albrecht2012obliquities,huber2013stellar,winn2017constraints,dai2017oblique,dalal2019nearly}. Large stellar obliquities were first widely observed among hot Jupiter hosts \citep{winn2010hot}, but have since been detected in multi-transiting systems \citep{huber2013stellar} and in systems with planets down to Earth's size \citep{kamiaka2019misaligned}. Therefore, non-zero stellar obliquities occur in many planetary systems at some point in their evolution; what is less certain is when these obliquities are excited.

Here, we have focused on scenarios whereby stellar obliquities are excited while the disk is still present, including gravitational perturbations from an exterior stellar companion \citep{batygin2012primordial,spalding2014early,lai2014star,zanazzi2018planet}, magnetic torques between the disk and star \citep{spalding2015magnetic,lai2011evolution} or simply side effects of star formation within a turbulent environment \citep{bate2010chaotic,spalding2014alignment,fielding2015turbulent}. However, it is currently unclear whether non-zero stellar obliquities typically emerge during these early times or through later, dynamical processes \citep{chatterjee2008dynamical,ngo2016friends,naoz2016eccentric,anderson2018teetering}. 

Unfortunately, most of the techniques leveraged to measure spin-orbit misalignments, such as the Rossiter-McLaughlin effect \citep{winn2005measurement}, gravity darkening \citep{barnes2009transit}, and asteroseimology \citep{huber2013stellar}, are rarely applicable to disk-hosting stars. An alternative approach is to photometrically constrain the stellar line broadening via $v\,\rm{sin}i_\star$ and divide by an estimated rotational velocity $v$ to constrain the inclination $i_\star$ \citep{winn2017constraints}. Among debris disks, this approach has suggested that stellar spin axes are usually aligned with the disk's plane to within a few 10s of degrees \citep{watson2011alignment,greaves2013alignment,matthews2013resolved}. 

More recently, \citet{davies2019star} extended similar investigations to protoplanetary disks, while noting that large debris disks may be less common among systems possessing star-disk misalignments. Of 15 disks studied, 5 exhibited misalignments between the stellar spin axis and the disk's plane exceeding $\sim10^\circ$. These measurements hint that primordial stellar obliquities are often small, but values of several 10s of degrees are relatively abundant. Moreover, systems that are truly misaligned can appear more aligned depending upon their sky-projected inclinations, potentially leading to an underestimation of their obliquities \citep{davies2019star}.

Reliable measurements of star-disk misalignments remain sparse, leaving the relationship between stellar spin axes and their disks poorly constrained. Nevertheless, our work here generates numerous predictions that may be tested using observed planetary systems. Specifically, if no exterior giant is present, we suggest that in the case of a slowly-dispersing disk, the inner planets adiabatically align with the stellar equator. This is consistent with the tendency thus far for multi-planet systems to exhibit low stellar obliquities \citep{sanchis2012alignment,2017AJ....154..270W,wang2018stellar,hirano2020evidence}. 

If, on the other hand, a distant giant is present during slow disk-dispersal, the inner planets first adiabatically align with the stellar equator, before returning to the giant's plane during subsequent stellar spin-down. The outcome is coplanarity among the inner planets and distant giant, but a potential obliquity for the host star. Recent observations have shown that distant giants are typically aligned with close-in multi-transiting systems \citep{masuda2020mutual}, but misaligned with transit singles.

The trends uncovered by \citet{masuda2020mutual} are consistent with both a giant-driven and stellar oblateness-driven origin to mutual inclinations. Thus, stellar obliquity emerges as a key observational feature that may partially disentangle these two mechanisms. Specifically, consider a system of close-in planets, coplanar with a distant giant, but misaligned with the stellar spin axis. Such a configuration can only exist if either; 1) the host star was never oblate enough to mutually incline the inner planets, or 2) the disk dispersed slowly. Stellar evolution models may in principle be used to rule out the former option, leaving the modern-day stellar obliquity as an empirical signpost of slow disk-dispersal in specific cases.

Unfortunately, measuring the stellar obliquity of close-in multi-planet systems that also possess a transiting exterior giant remains a significant observational challenge. The post-main sequence star Kepler-56 possesses a giant companion of $m_G\,\textrm{sin}\,I=5.6M_J$ and $a_G=2.2$\,AU (corresponding to $m_G/a_G^3\approx 0.5 M_J/\textrm{AU}^3$; \citealt{huber2013stellar,otor2016orbit}; also see Figure~\ref{Giant_Compare}), orbiting exterior to 2 transiting planets that are inclined with respect to the stellar spin axis by over $40^\circ$. The exterior giant's orientation is unknown, but if shown to transit, it may support the mechanism described here.

In short, spin-orbit misalignments during the disk-hosting stage are expected on theoretical grounds \citep{batygin2012primordial,spalding2015magnetic,lai2014star,fielding2015turbulent}. Observational confirmation of this expectation has arisen for a limited number of cases \citep{davies2019star}, but a substantially larger set of measurements is required in order to deduce a statistical distribution of star-disk misalignments. \\

\subsection{Dependence upon stellar type}

In our analysis, we restricted attention to the case of a star with Solar mass and radius. However, pre-main sequence stars contract from large stellar radii of $>2R_\odot$ to a final value that depends upon the stellar mass. Moreover, this contraction occurs over a timescale comparable to the disk lifetime. Upon leaving their Hayashi tracks, stars with $M_{\star} \gtrsim 0.3M_\odot$ develop a radiative core, affecting $k_2$  \citep{gregory2016influence}. Consequently, the stellar quadrupole weakens substantially during pre-main sequence evolution. 

In turn, disk lifetimes vary widely. Shorter-lived disks will disperse while the stellar quadrupole is stronger, as compared to longer-lived disks. More massive stars appear to lose their disks sooner \citep{ribas2015protoplanetary}, and arrive onto the main sequence with larger radii \citep{gregory2016influence}.  Consequently, we expect a modest trend whereby higher-mass stars drive greater mutual inclinations among their close-in planets. A factor confounding this expectation is that a stronger stellar quadrupole augments the secular frequencies, such that faster disk dispersal timescales would be required to excite mutual inclinations. As discussed below, more information regarding disk dispersal timescales is required to delineate this expectation.

In contrast to radii, stellar spin rates are remarkably uniform across time and stellar type during the disk-hosting stage \citep{bouvier2014angular}. However, over longer timescales, stars with $M_{\star} \gtrsim 1.2 \ M_\odot$ tend to retain rapid rotations throughout their entire main-sequence lifetimes \citep{kraft1967studies,skumanich1972time}. Combined with larger radii, these more massive stars exert a strong quadrupolar influence upon close-in planets long after disk dispersal, preventing adiabatic alignment with an exterior giant. Accordingly, we expect stars with $M_\star \gtrsim 1.2 \ M_\odot$ hosting multi-planet systems to exhibit smaller stellar obliquities, even if a distant giant is known to exist.  

\subsection{Disk dispersal timescale}
\label{sec: Timescale}

In previous investigations of star-driven misalignment, disk dispersal was assumed to be instantaneous \citep{spalding2016spin,spalding2018resilience,li2020mutual}. Here, we have shown that this assumption is invalid unless the disk's gravity vanishes over a timescale comparable to or shorter than the system's slowest secular eigenmodes, typically between $10^{2-4}$ years, with timescales below $10^3$\,years required for fiducial parameters (Figure~\ref{EigenSlow}). If disks typically disperse more slowly than this limit, then in general mutual inclinations among close-in planets are more likely to arise through later dynamical mechanisms that incline exterior giants \citep{hansen2017perturbation,2017AJ....153...42L,2018MNRAS.478..197P,gratia2017outer}.

It is important to emphasize that the ``disk dispersal timescale" mentioned here is distinct from the ``disk lifetime". Disk lifetimes are relatively well constrained to lie between $\sim1-10\,$Myr \citep{haisch2001disk,armitage2011dynamics,alexander2014dispersal,mamajek2009initial,ribas2015protoplanetary}, with occasional longer-lived outliers \citep{silverberg2020peter}. On the other hand, despite their significance, disk dispersal timescales are poorly constrained. Disks undoubtedly vanish over a timescale that is much shorter than their multi-Myr lifetimes--a conclusion apparent from a lack of examples of disks that are in the process of dispersing \citep{cieza2008masses,koepferl2013disc,alexander2014dispersal}. Moreover, the current inclinations of asteroids require that the inner several AU of the Solar nebula must have dispersed more rapidly than $\sim 10^{4-5}$\,years \citep{Ward1981solar}. Beyond these crude upper limits, little empirical data exists to constrain typical dispersal timescales.

Theoretical estimates of disk dispersal times are hampered by uncertainties regarding the dominant mechanism(s) driving their dispersal. A leading hypothesis is that UV photoevaporation of disk material at a few AU drives a wind of material from the disk's surface \citep{alexander2014dispersal}. As the disk ages and its accretion slows, the photoevaporative wind eventually exceeds the rate of inward disk-driven accretion. At this point, the inner few AU of the disk is cut off from resupply and viscously drains onto the star. Modeling the photoevaporation process predicts that the inner 0.2\,AU can drain on timescales of $\sim 10$\,kyr \citep{alexander2006photoevaporation,owen2010radiation,gorti2015impact}.

A simple expression for the viscous time in terms of the dimensionless parameter $\alpha\sim 10^{-3}$ may be written as \citep{shakura1973black}
\begin{align}\label{viscous}
\tau_\nu\sim\frac{a^2}{\alpha\Omega h^2}\approx 5700 \bigg(\frac{a}{0.2\,\textrm{AU}}\bigg)^{3/2}\,\textrm{years}.
\end{align}
Though only a crude approximation, this timescale suggests that once starved from replenishment, the inner disk may disperse on timescales comparable to millennia. This is more rapid than the adiabatic threshold for small $J_2$ values (Figure~\ref{EigenSlow}), but may prevent mutual inclinations from being excited in systems with stronger $J_2$ and larger eigenvalues. Thus, paradoxically, larger stellar quadrupoles may inhibit mutual inclinations by increasing the eigenvalues and allowing planetary orbits to adiabatically reorient during disk dispersal.

In our simulations, we simply assumed that the disk's surface density decayed homogeneously, and the disk remained coplanar throughout. However, this assumption is only valid if the inner disk mass is larger than that of the planets. We may solve for the disk mass required to dominate the quadrupolar potential felt by the planets by setting $\nu_{\star,i}=\nu_d$. This requirement yields a disk scaling factor of
\begin{align}
\Sigma_{0,\mathrm{crit}}=\frac{3}{2}J_2 \bigg(\frac{a_1}{a_0}\bigg)\bigg(\frac{R_\star}{a_1}\bigg)^2\frac{\beta M_\star}{\pi a_1^2}
\end{align} 
The mass interior to radius $a_{out}$ at this surface density may be computed as
\begin{align}
M_{\mathrm{int}}&=\int^{a_{\mathrm{out}}}_02\pi a \Sigma_{0,\mathrm{crit}}\bigg(\frac{a_0}{a}\bigg) da\nonumber\\
&=3J_2 \bigg(\frac{R_\star}{a_1}\bigg)^2\bigg(\frac{a_{\mathrm{out}}}{a_1}\bigg)\beta M_\star\nonumber\\
&\approx \bigg(\frac{J_2}{10^{-3}}\bigg)\bigg(\frac{a_1}{0.1\textrm{AU}}\bigg)^{-3}\bigg(\frac{M_{\star}}{M_\odot}\bigg) \bigg(\frac{a_{\mathrm{out}}}{\textrm{AU}}\bigg)M_\oplus.
\end{align}
This result suggests that the disk's influence is comparable to the star's right down to $M_{\mathrm{int}}\sim M_{\oplus}$. This is contradictory, since the planets in our simulations are 5 times that mass, violating the assumption that the disk remains coplanar. Moreover, the importance of disk dispersal time extends beyond the super-Earth regime considered here and into the regime of hot Jupiter inclinations \citep{zanazzi2018planet}, a mass regime long considered to carve out gaps in the natal disk \citep{goldreich1980disk,zhu2011transitional}. Accordingly, close-in planets undoubtedly contribute to the final disruption of the natal disk, and may even hasten disk dispersal. The importance of planet-disk interactions for disk dispersal timescales is therefore an important outstanding problem that merits further study.

\section{Summary}
\label{sec: summary}

 In this paper, we explored two different but complementary hypotheses for the generation of mutual inclinations within close-in multi-planetary systems. In the first hypothesis, the quadrupolar moment arising from an inclined, rapidly-rotating star drives mutual inclinations soon after the dispersal of the protoplanetary disk \citep{spalding2016spin,spalding2018resilience,li2020mutual}. The second hypothesis holds that the secular perturbation arising from an inclined giant planet, orbiting exterior to the close-in system, dynamically excites mutual inclinations \citep{hansen2017perturbation,2017AJ....153...42L,2018MNRAS.478..197P}. Previous work had not compared the relative importance of each of these hypotheses. 
 
 Here, we showed that at the early stage of disk dispersal, the quadrupolar potential of the host star typically exceeds the secular potential of an exterior giant planet. Moreover, distant giants are expected to form coplanar with their interior planets. They therefore require dynamical interactions to occur after disk dispersal in order to acquire mutual inclinations. In contrast, stellar obliquities may arise during the disk-hosting stage by a variety of mechanisms. Accordingly, the stellar quadrupole is likely to play a greater role in generating mutual inclinations among close-in planets immediately following disk-dispersal, as compared to distant giants (see Figures~\ref{2p} \& \ref{3p}). 

Our most critical findings are summarized in Figure~\ref{schematic}. Specifically, while the protoplanetary disk is present, the inner planets are likely coplanar with a distant giant, while the star may possess a non-zero obliquity. Earlier versions of the hypothesis had proposed that the disk disperses instantly, such that the planets exactly inherit the plane they possessed during the disk-hosting stage \citep{spalding2016spin,spalding2018resilience,li2020mutual}. In contrast, we show that the disk's gravitational influence must vanish over a timescale that is significantly shorter than roughly  $10^{3}$ years in order for the planets to approximately retain the disk's orientation. This timescale is substantially shorter than disk lifetimes of $10^{6-7}$ years, but comparable to the viscous timescale over which the inner $0.2\,$AU of the disk gas may disperse.

If, on the other hand, the disk disperses over timescales longer than several thousand years, the inner planets adiabatically reorient towards the stellar spin axis, exciting negligible mutual orbital inclinations. Subsequently, as the star spins down and loses its quadrupolar moment, the inner planets reorient towards the distant giant's plane. We predict that this more gradual process leads to a multi-transiting system, coplanar with the distant giant, exhibiting a stellar obliquity if one existed at the disk-hosting stage. Stars with $M_{\star}\gtrsim1.2\,M_\odot$ tend to retain their rapid rotations, thereby maintaining relative inclinations between their outer giants and interior planetary systems. We hope that these features may become observable within upcoming surveys. 

Disk-dispersal is a poorly-understood process, and its associated timescales are largely unconstrained. Whereas disks live for millions of years, their removal is relatively brief, particularly in the inner regions. Our work here highlights the critical importance of disk-dispersal for sculpting the final architectures of planetary systems, with implications even for giant planets residing at multiple AU. In this view, planetary systems can only be understood in a holistic sense; the star is inextricably linked, via the inner planets and natal disk, to the outermost reaches of the planetary system. 

\section{Acknowledgements}
We thank Kento Masuda for stimulating conversations and Konstantin Batygin for useful comments. We are also grateful for the thoughtful input of an anonymous reviewer whose comments significantly improved the narrative and content of the paper. C.S. thanks the 51 Pegasi b Heising-Simons Foundation grant for their generous support. S.M. was supported by the NSF Graduate Research Fellowship Program under Grant DGE-1122492. Support for this work was also provided by NASA through the NASA Hubble Fellowship grant \#HST-HF2-51465 awarded by the Space Telescope Science Institute, which is operated by the Association of Universities for Research in Astronomy, Inc., for NASA, under contract NAS5-26555. This research has made use of the NASA Exoplanet Archive, which is operated by the California Institute of Technology, under contract with the National Aeronautics and Space Administration under the Exoplanet Exploration Program.


\end{document}